\documentclass{sig-alternate-05-2015}
\usepackage{graphicx,color,url,footmisc}
\usepackage{booktabs}
\usepackage{synttree}
\usepackage{multirow}
\usepackage{amsmath} 
\usepackage{caption}

\usepackage{subfig}
\usepackage{balance}

 \newcommand{\myparatight}[1]{\smallskip\noindent{\bf {#1}:}~}

\newenvironment{packeditemize}{\begin{list}{$\bullet$}{\setlength{\itemsep}{2pt}\addtolength{\labelwidth}{4pt}\setlength{\leftmargin}{20pt}\setlength{\listparindent}{\parindent}\setlength{\parsep}{0pt}\setlength{\topsep}{0pt}}}{\end{list}}

\begin{document}

 \CopyrightYear{2016} 
\setcopyright{acmcopyright}
\conferenceinfo{ASIA CCS '16,}{May 30-June 03, 2016, Xi'an, China}
\isbn{978-1-4503-4233-9/16/05}\acmPrice{\$15.00}
\doi{http://dx.doi.org/10.1145/2897845.2897908}

\title{Forgery-Resistant Touch-based  Authentication on Mobile Devices}

\numberofauthors{4}

\author{
\alignauthor Neil Zhenqiang Gong\\
       \affaddr{ECE, Iowa State University}\\
       \email{neilgong@iastate.edu}
\alignauthor Mathias Payer\\
       \affaddr{CS, Purdue University}
       \email{mathias.payer@nebelwelt.net}
\and
\alignauthor Reza Moazzezi\\
       \affaddr{EECS, UC Berkeley}\\
       \email{rezamoazzezi@berkeley.edu}
\alignauthor  Mario Frank\\
       \affaddr{EECS, UC Berkeley}
       \email{mail2mf@gmx.de}
}
\maketitle

\begin{abstract}
Mobile devices store a diverse set of private user data and have gradually
become a hub to control users' other personal Internet-of-Things devices.
Access control on mobile devices is therefore highly important.  The widely
accepted solution is to protect access by asking for a password.  However,
password authentication is tedious, e.g., a user needs to input a password every
time she wants to use the device. Moreover, existing biometrics such as face,
fingerprint, and touch behaviors are vulnerable to forgery attacks. 

We propose a new touch-based biometric authentication system that is passive and
secure against forgery attacks.  In our touch-based authentication, a user's
touch behaviors are a function of some random ``secret''.  The user can
subconsciously know the secret while touching the device's screen. However, an
attacker cannot know the secret at the time of attack, which makes it
challenging to perform forgery attacks even if the attacker has already obtained
the user's touch behaviors.  We evaluate our touch-based authentication system
by collecting data from 25 subjects.  Results are promising: the random secrets
do not influence user experience and, for targeted forgery attacks, our system
achieves 0.18 smaller Equal Error Rates (EERs) than previous touch-based
authentication.  

\end{abstract}


%
%

%
%



\section{Introduction}

Since the introduction of the first iPhone by Apple in June 2007, touch based
mobile devices have become ubiquitous. For instance,  the volume of the
smartphone market has already surpassed that of the PC market in
2011~\cite{smartphoneMorethanPC}. Users often store a large amount of sensitive
data on mobile devices. Moreover, with the advent of Internet-of-Things (IoT)
devices such as smartwatches, fitness trackers, the Nest Thermostat~\cite{Nest},
and medical devices like Bee~\cite{Bee}, smartphones have gradually become the
hub of IoT. Specifically, a user could use a smartphone to control her
smartwatch, adjust home temperature via remotely controlling the Nest
Thermostat, and view the user's insulin injection data and glucose levels from
Bee. Access control on mobile devices is important because having access to a
user's mobile device allows an attacker to 1) access the user's personal data,
and 2) control the user's other connected IoT devices and access sensitive data
on them. For instance, an attacker that obtains access to a smartphone can
access the user's sensitive SMS messages, emails, and apps, 
as well as manipulate the user's home temperature by remotely
controlling the connected Nest Thermostat.

 The most popular method to address such threats is to authenticate a user via
\emph{password} before allowing her to use the device, i.e., the user logs in
the device with a correct password. However,  users might turn off password
authentication because it is tedious and
inconvenient~\cite{NolockSOUPS13,whylockCCS14,sophos}. For instance,  Egelman et
al.~\cite{whylockCCS14} showed that 42\% of users do not lock their smartphones,
and 34\% of them do so because locking is ``too much of a hassle". Moreover, it
is well known that  conventional biometrics such as face, fingerprint, and voice
are  vulnerable to forgery
attacks~\cite{AttackingAndroidFaceAuthentication,AttackingAndroidFaceAuthenticationLiveness,fingerprintForgeryAttack}.
For instance, fingerprint readers can be tricked by taking an image of the
fingerprint, forming a mold, and using wood glue to make a fake
finger~\cite{fingerprintForgeryAttack}. Therefore, it is urgent to design secure
and usable authentication methods. 
 







A number of recent studies have raised the possibility of using low level
interactions such as how a user touches the screen  as a biometric signature for authentication in
mobile devices~\cite{bo2013silentsense, munichGuys, frank2013touchalytics,
li2013unobservable, sae2012investigating}.  The key idea for such authentication
mechanisms is that users produce touch data all the time when using the device so
that authentication can be \emph{passive}, i.e., without requiring  the user to
carry out any action dedicated to authentication. 


However, this touch-based authentication mechanism, like conventional
biometrics, is also vulnerable to forgery attacks. For instance, an attacker
(e.g., a `friend' or the spouse of the targeted user) can collect the targeted
user's touch data via convincing her to use the attacker's mobile devices and
recording her touch data. Later, the attacker can program a Lego robot to replay
the collected touch data on the targeted mobile device, which can compromise the
authentication system with  a high probability~\cite{attack-CCS13}.  


\myparatight{Our work} In this work, we demonstrate a defense against forgery
attacks to touch-based biometrics. 
In particular, we  defend against forgery attacks by leveraging the impact of
screen settings (serving as a random ``secret") on a user's touch behavior. The
sensors on the screen of a mobile device record where, when, how fast, and how
heavily a user's finger touches the screen.  Before the recorded raw data is
sent to applications on the mobile device, our approach transforms the data as
it passes through the operating system according to a \emph{screen setting}.
For instance, a screen setting of 0.8 horizontal distortion means that a 1.0cm
long horizontal line starting at a certain location  on the screen is received
by the application as a 0.8cm long horizontal line starting at the same
location. Due to such modifications, the running applications react differently
to the actions of the user. As a consequence, the user will adapt her touch
behavior (i.e., raw touch data recorded by screen sensors) in order to achieve
the desired application behavior.  Ideally, the adaptation is performed
subconsciously, i.e., the user does not explicitly notice that the screen
settings have changed but still adapts the touch behavior to compensate for this
change. We investigate the impact of screen settings on a user's touch behavior
and their applications to defend against forgery attacks. 

\begin{figure}[!t]
\centering
{\includegraphics[width=0.40\textwidth]{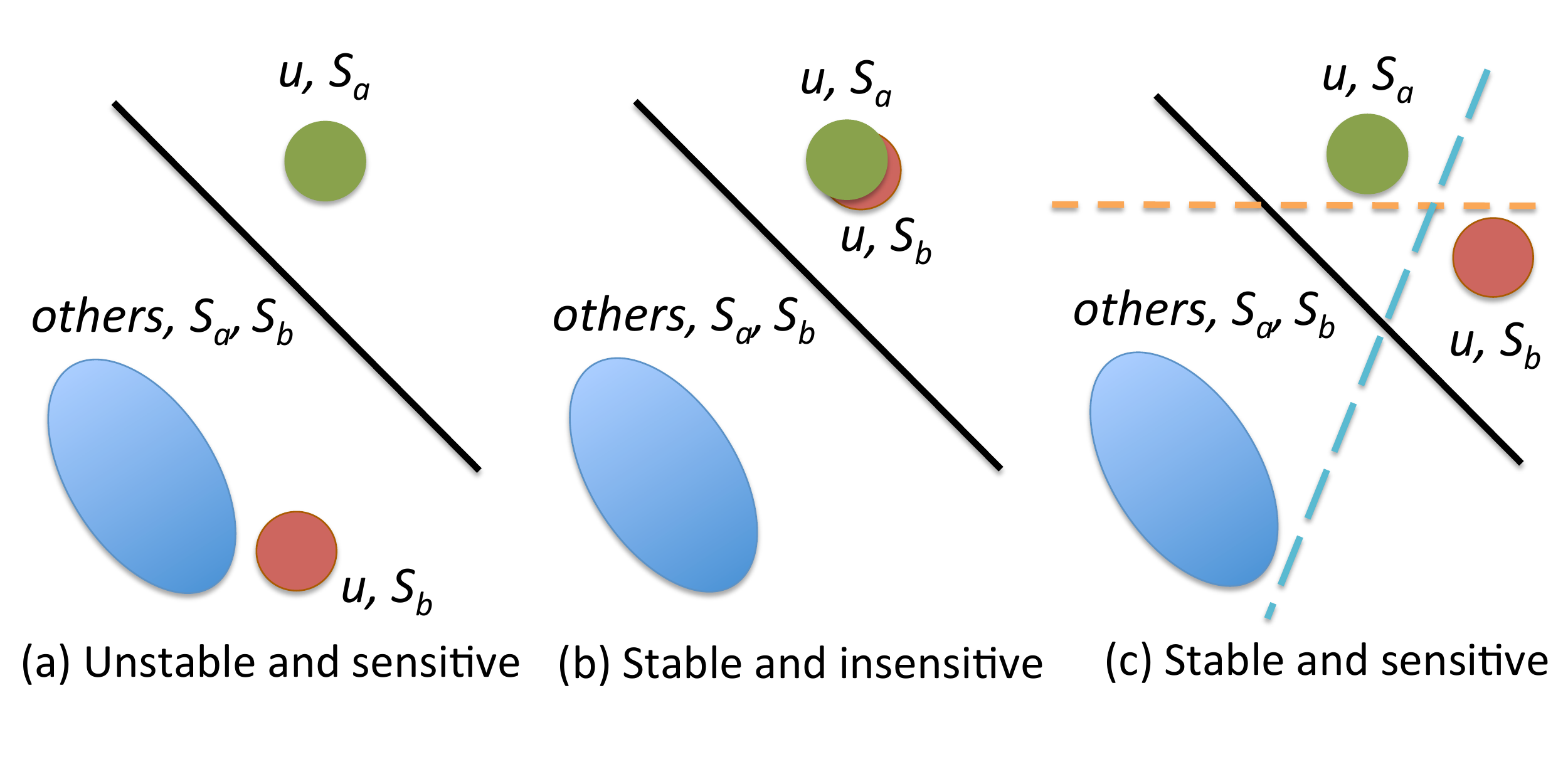}}
\vspace{-5mm}
\caption{Concepts of stability and sensitivity.}
\label{sta-sen}
\vspace{-3mm}
\end{figure}

First, we find that, when screen settings are discretized properly,
  a user's touch behavior  in two different screen settings is \emph{stable},
 meaning that the behavior difference between two different users in the same setting 
is larger than that between two different settings for the same user. 
Stability guarantees that we can distinguish 
a user's touch behavior from other users'. 
Unfortunately, stability implies that if we learn a model to distinguish
 a user's touch behavior from other users' using only a single setting, 
then forgery attacks that replay the targeted user's data collected
 in other settings succeed with a high probability.  
Extending stability, we find that a user's touch behavior in different screen
 settings is also \emph{sensitive}, meaning that they have a high 
degree of separability in the feature space.
 Sensitivity guarantees that one can learn a model to distinguish
 touch behavior of a user in two different screen settings,
 serving as a basis for a sufficient defense against forgery attacks. Figure~\ref{sta-sen}
 shows the concepts of stability and sensitivity.


Second, based on our findings, we propose a novel  continuous authentication
mechanism called \emph{adaptive touch-based continuous authentication}. Our
system consists of a \emph{registration phase} and an \emph{authentication
phase}. In the registration phase, we sample a set of screen settings in which a
user's touch behavior is both stable and sensitive. Then we train a model for
each setting $s$ to distinguish the behavior of the user in $s$  from the
behavior of the same user in other settings and those of other users in all
considered settings.  In the authentication phase, we randomly sample a
predefined setting and use the corresponding model to continuously authenticate
the user in each time interval. Our authentication system can significantly
decrease the success rate of an attacker who knows the targeted user's touch
data in \emph{all} settings. This is because the attacker cannot know the
setting used by our system at the time of attacks and replaying touch data
collected in a different setting fails to pass the authentication with high
probability.  

Third, we evaluate our system via collecting data from 25 subjects in five
different settings along the X axis and five different settings along the Y
axis.  We find that users can subconsciously adapt their touch behavior to
different screen settings, i.e., the transitions between two settings do not
interrupt  users nor affect user experiences. 
Moreover, our system achieves 0.02 to 0.09 smaller  mean Equal Error Rates
(EER) than previous work
for random forgery attacks and 0.17 to 0.18 smaller mean EERs than previous work
for targeted forgery attacks; the  registration phase of our authentication
system takes a short period of time, i.e., touch data collected within two
minutes  are enough to train a model for a setting; and our system achieves
smaller EERs with more screen settings.

In summary, our key contributions are as follows:
\begin{packeditemize}
  \item We demonstrate the stability and sensitivity of touch behavior to screen
settings.
	\item We propose a new touch-based continuous authentication mechanism, which
builds on the stability and sensitivity properties of behavioral touch patterns
to achieve forgery-resistant touch signatures. 
	\item We evaluate our authentication system via collecting touch data from
 25 subjects in five  screen settings along the X axis and five screen settings
along the Y axis. We demonstrate that our system significantly outperforms
previous work at defending against forgery attacks. 
\end{packeditemize}

\section{Background and Related Work} \myparatight{Screen settings}
Users  interact with a mobile device via swiping, clicking, or zooming on the
screen. For instance, users often slide over the screen horizontally (e.g.,
navigate to the next page of icons in the main screen or browse through photos)
and vertically (e.g., read webpages, social media updates, or emails), which
result in horizontal swipes and vertical swipes, respectively. A swipe is also
called a \emph{stroke}, and we will use them interchangeably in the paper. 

The sensors on the screen record the location, timing, pressure, and covering
area of interactions.  More specifically, an interaction is a sequence of touch
points ($t_i$,$x_i$,$y_i$,$p_i$,$a_i$), where $t_i$ is the timestamp, $x_i$ is
the horizontal location, $y_i$ is the vertical location,  $p_i$ is the pressure,
and $a_i$ is the covering area  of the $i$th touch point.   Then these touch
points are transformed by the operating system to higher-level primitives such
as clicks, swipes, and zooms.  Applications on the mobile device access the
transformed primitives through the operating system.  In the following,
\emph{touch behavior} refers to the raw touch data recorded by the screen
sensors. 

A \emph{screen setting}  controls how the sensed raw data is transformed to the
primitives that are used by applications. For instance, screen settings can
independently distort the X (i.e., horizontal) axis or the Y (i.e., vertical)
axis. 
Note that  transformation is only applied to a sequence of touch points and the
first touch point of any interaction is not transformed. Therefore, one-touch
operations like clicking a button on an application will not be affected by
different screen settings.  Moreover, since screen settings are transparent to
applications, application developers do not need to change their applications
when the screen settings are modified.

Suppose a user draws a line from position (10,10) to position (110,110) on the
screen.  Under a screen setting of 0.8 Y-distortion the application would see a
line from position (10,10) to position (110,90). Under a screen setting of 1.2
X-distortion the application would see a line from position (10, 10) to position
(130, 110). To account for such transformations, the user will adapt her touch
behavior  to achieve the desired application behavior.

\myparatight{Touch-based continuous authentication}  Touch-based authentication
has been proposed first in \cite{munichGuys} and
\cite{Sae-Bae:2012:BGN:2207676.2208543}. However, both methods require the user
to carry out a specific secret gesture at a defined point of time (unlock
challenge) and then analyze \emph{how} the gesture is carried out. More
recently,  leveraging  a user's  touch interactions obtained when the user
interacts with the device to continuously and implicitly monitor the user has
attracted increasing attentions. 
Complex interactions such as zooming are too infrequently used to be appropriate
for continuous monitoring and clicks exhibit too few  features to be
discriminative for users. 
 Therefore, most previous work focus on  swiping interactions (i.e., strokes),
which were demonstrated to contain a behavioral biometric signature that may be
used to continuously authenticate the user~\cite{bo2013silentsense,
frank2013touchalytics, li2013unobservable, sae2012investigating,
sherman2014user}. 

For instance, Frank et al.~\cite{frank2013touchalytics} extracted 31 features
from each stroke and trained a classifier for a user to distinguish her touch
behavior from other users'. These features include the direction of the
end-to-end line, average velocity, start locations, and end locations of  a
stroke. For a complete list of the features, please refer to Frank et
al.~\cite{frank2013touchalytics}.
More recently, Sae-Bae et al.~\cite{sae2014multitouch} studied a canonical set
of 22 multitouch gestures for authentication on mobile devices, and they found a
desirable alignment of usability and security, i.e., gestures that are more
secure are rated by users as more usable. Sherman et al.~\cite{sherman2014user}
proposed to use free gestures as a static authenticator to unlock  mobile
devices and they further studied the memorability of user generated gestures.
Xu et al.~\cite{Xu2014soups}  verified that touch-based authentication is a
promising  authenticator  via conducting experiments with around 30 users for
one month in the wild.

%
 

However, all these touch-based authentication mechanisms consider a \emph{single} 
universal screen setting (e.g., the default screen setting) for all users, 
making them vulnerable to \emph{forgery attacks} which collect a 
user's touch strokes in the screen setting and program a robot to 
replay them to attack the authentication system 
(please see more details in Section~\ref{sec:exp}). 
Our work also focuses on users' strokes, but we leverage multiple, randomly
chosen screen settings.



\myparatight{Common-behavior attacks} Serwadda and Phoha~\cite{attack-CCS13}
showed that a Lego robot can be programmed to swipe the screen of a mobile
device, and the stroke recorded by the screen sensors is as desired. 
The attacker needs to
know the defining parameters (e.g., start and stop
locations) of the stroke that is to be forged. 

Moreover,  they proposed that the attacker can simply program the robot to
replay the common touch behavior of a large population onto the targeted user's
mobile device, and they showed that such \emph{common-behavior attacks} can
significantly increase the EERs. However, they also showed that their
common-behavior attacks have limited success rates for users whose behaviors are
relatively far from the common, and there are about 20-40\% of such users. This
means that touch-based authentication is appropriate to these users. In
practice,  touch-based authentication systems can compare a user's touch
behavior to those of a large population and recommend if the behavior is far
enough from the common behaviors so that it is resilient to common-behavior
attacks. 

In this work, we consider stronger forgery attacks in which the attacker
 could obtain the targeted user's touch data.
 
 

\section{Threat Model}
\label{sec:threat}

The authentication system is available to the attacker. The attacker can
read and analyze the implementation details of the authentication system
offline. Therefore, if the authentication system uses a universal screen setting
(e.g., the default screen setting) for all users,
the attacker can obtain this setting via offline code analysis. However, we
assume that the attacker cannot obtain the dynamic behaviors of the
authentication system that runs on the targeted user's mobile device. For instance, 
the attacker cannot know the
current setting used by the authentication system if it is randomly sampled  in
random time intervals. This is because 
reading out the current settings at runtime requires high
privileges (e.g., root access to the operating system) and an attacker that has
already obtained such high privileges already compromised the system.

 We assume the attacker has a commercialized programmable 
 robot (e.g., a Lego robot), 
 which can be used to forge touch strokes and play them on mobile devices. 
 For instance, Serwadda and Phoha~\cite{attack-CCS13} showed 
 how to program a Lego robot to forge  touch strokes  to have desired  features. 
 We note that an intelligent robot which is equipped with specialized sensors could potentially detect the current screen setting used by our authentication system using advanced computer vision algorithms. However, the attacker might not have such a specialized robot, and thus, in this work, we consider  \emph{state-of-the-art commercialized} robots, with which the attacker cannot infer the current screen setting.

\begin{figure*}[!t]
\centering
\subfloat[User A, 0.8 Y-distortion]{\includegraphics[width=0.16\textwidth, height=1.6in]{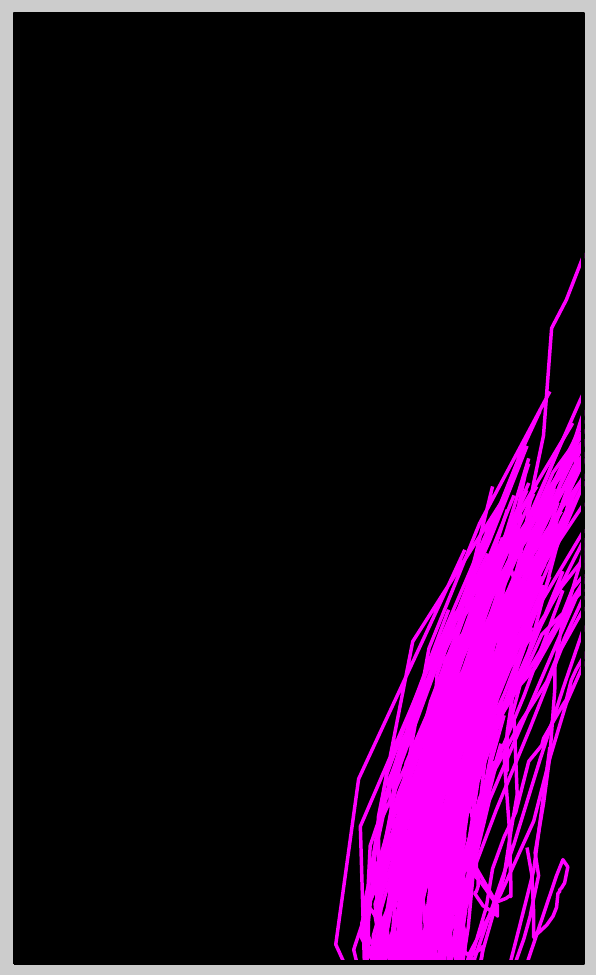}}
\hspace{8mm}
\subfloat[User B, 0.8 Y-distortion]{\includegraphics[width=0.16\textwidth, height=1.6in]{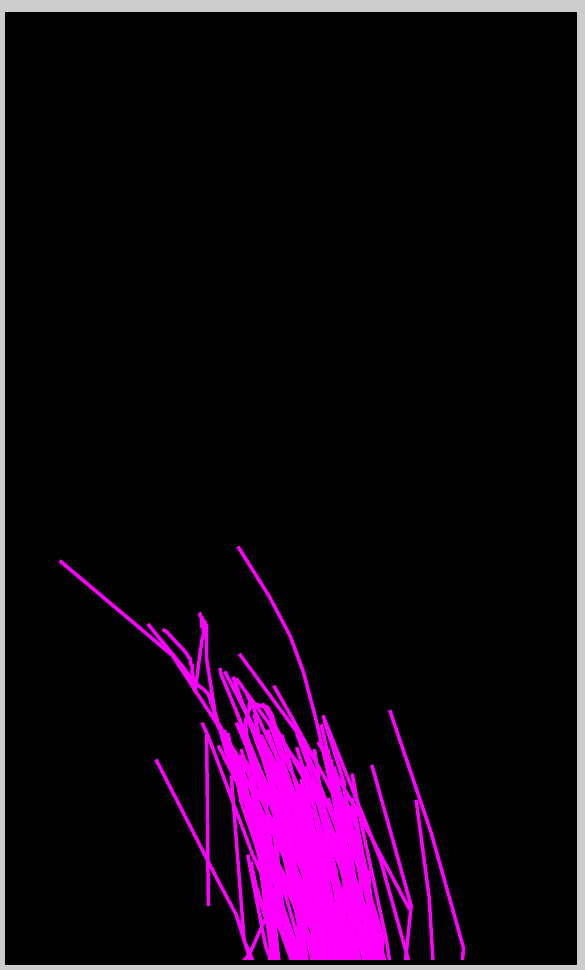}}
\hspace{8mm}
\subfloat[User C, 0.8 Y-distortion]{\includegraphics[width=0.16\textwidth, height=1.6in]{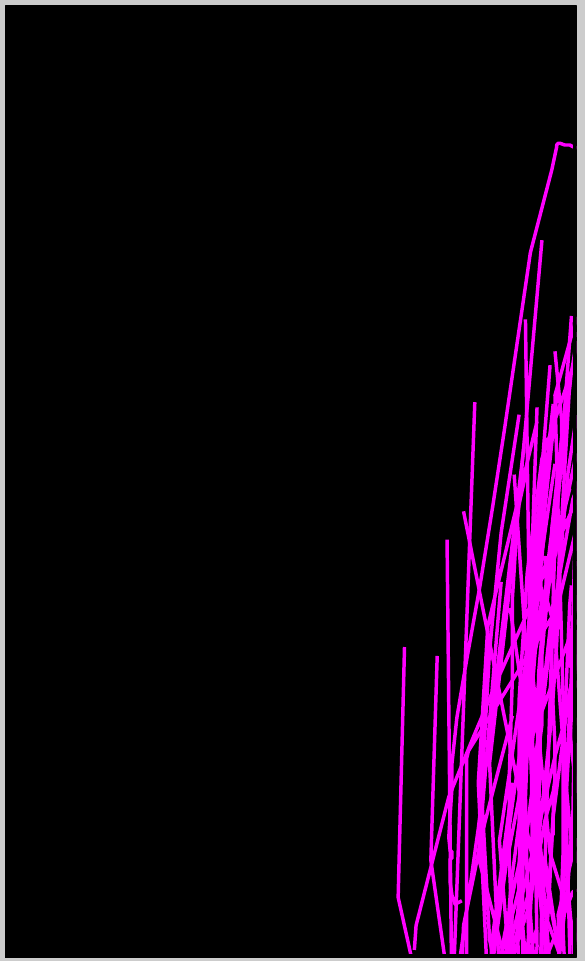}}

\subfloat[User A,  1.2 Y-distortion]{\includegraphics[width=0.16\textwidth, height=1.6in]{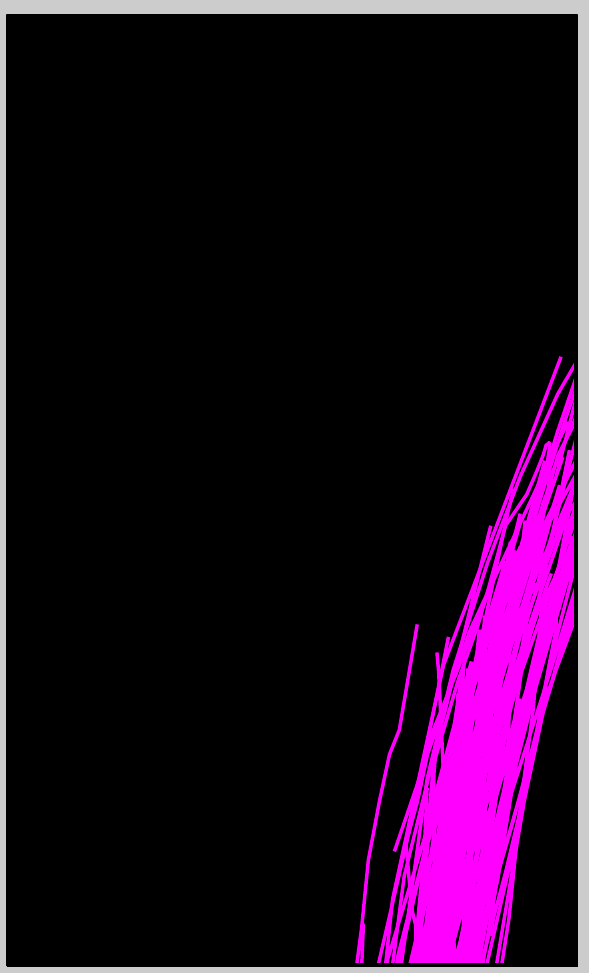}}
\hspace{8mm}
\subfloat[User B,  1.2 Y-distortion]{\includegraphics[width=0.16\textwidth, height=1.6in]{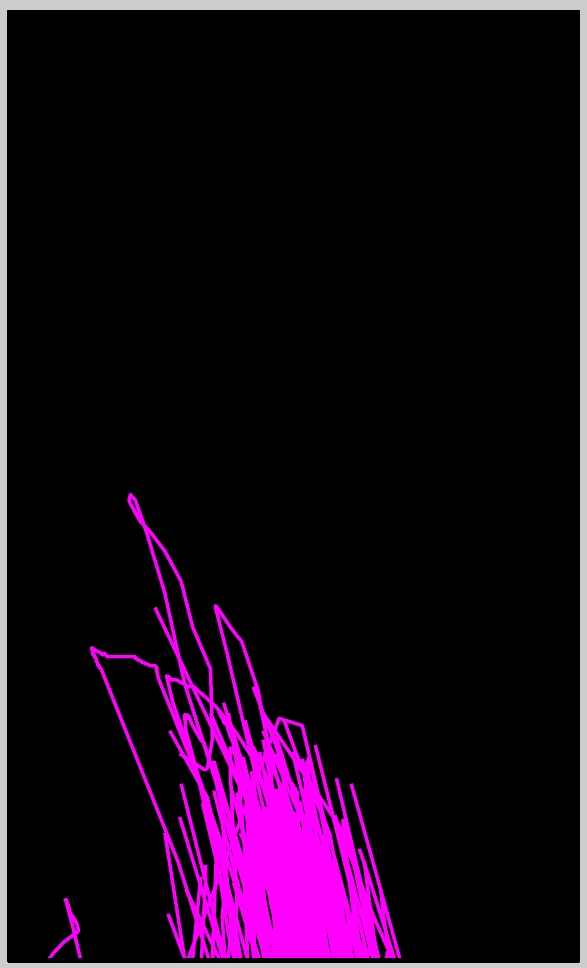}}
\hspace{8mm}
\subfloat[User C, 1.2 Y-distortion]{\includegraphics[width=0.16\textwidth, height=1.6in]{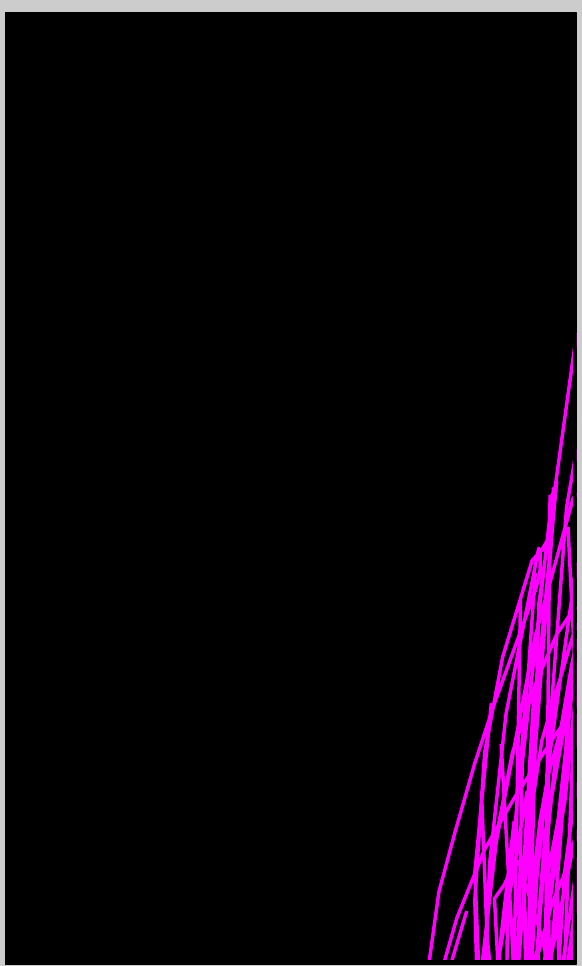}}
\caption{Vertical strokes of three users A, B, and C in two screen settings, which we obtained in our experiments. 
The black background simulates the mobile device's screen while the lines are users' touch strokes.  We observe that users' touch
 behaviors are \emph{stable}, i.e., the touch behaviors of a user in one setting
are closer to those of the user in another setting than those of other users.}
\label{stroke} 
\end{figure*}

We consider two general forgery attacks depending on whether the attacker obtains touch
strokes of the targeted user or not. 
 We will discuss more advanced targeted attacks such as training a human attacker in Section~\ref{discussion}.

\myparatight{Random attacks} In this scenario, the attacker does not obtain
touch strokes of the targeted user. For instance, this could be the case in
which the targeted user lost the device and it is found by a random
attacker, who does not know the targeted user and does not have its touch
strokes.  However, the attacker could obtain touch strokes of a set of other
users. This is reasonable because 1) the attacker can retrieve touch strokes
from publicly accessible data sets~\cite{frank2013touchalytics,attack-CCS13},
and 2) it is possible that users (intentionally or unintentionally) install  an
application (e.g., this application is a fun game application and  does not
appear to be malicious) that is developed by the attacker on their mobile devices 
and the application records users' touch strokes.  

After obtaining touch strokes from a set of users in different screen settings,
the attacker randomly selects touch strokes, programs a Lego robot 
to forge them~\cite{attack-CCS13}, and uses the programmed robot to 
touch the mobile device.

\myparatight{Targeted attacks} In this case, the attacker obtains touch strokes
of the targeted user. For instance, the attacker could be  a ``friend'' of the
targeted user or the targeted user's curious spouse, who wants to access the
messages sent by the targeted user or know whom the targeted user has called,
and the attacker   convinces the targeted user to use his/her mobile devices to
record the targeted user's  touch strokes. 

Again, the attacker programs a  robot using the collected strokes to attack the targeted user.
Intuitively, if the authentication system uses an universal screen setting, the attacker 
can obtain the targeted user's strokes in this universal setting, which results in attacks 
with very high success rates (see Section~\ref{sec:exp}). Our new authentication mechanism uses multiple screen settings 
in which a user's touch behavior is different, and we randomly sample one of the screen settings 
in each time interval. 
Given enough settings, our system can significantly decrease the success rates of targeted attacks even if the attacker 
obtains the targeted user's  touch strokes 
in \emph{all} screen settings. This is because the attacker cannot know the screen setting used by our
 authentication system at the time of attack, and thus the attacker does not know which strokes should be used to
program the robot. Therefore, the attack is reduced to randomly guessing a setting $s$, but replaying strokes collected in $s$  
 passes the authentication with much lower probabilities if $s$ is not the current setting used by
 our authentication system. 
 We note that we do not consider attacks with advanced robots to automatically infer the screen setting in this work.

%
%


\section{Stability and Sensitivity}


\begin{figure*}[!t]
\centering
\subfloat[User A]{\includegraphics[width=0.33\textwidth]{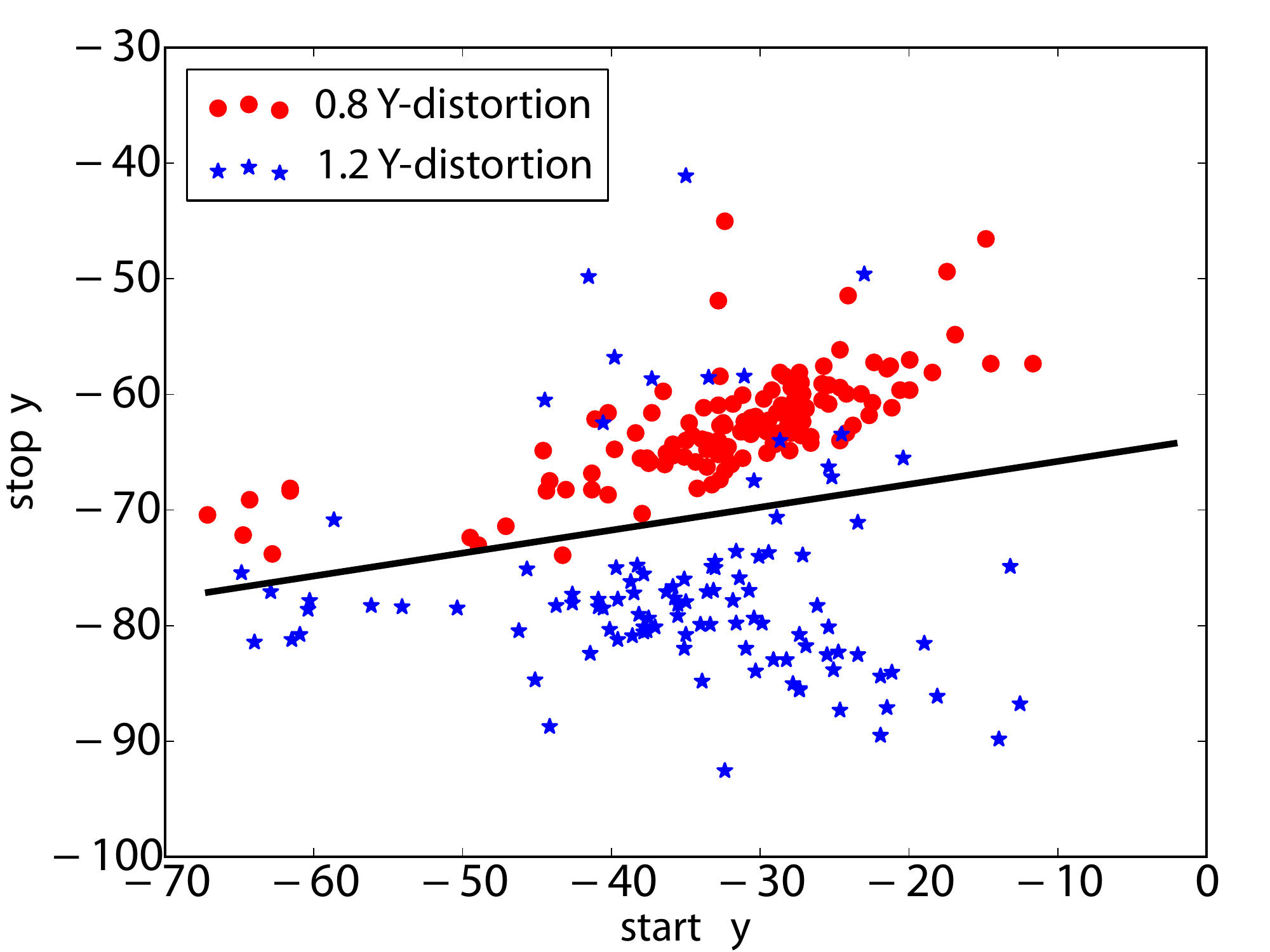}}
\subfloat[User B]{\includegraphics[width=0.33\textwidth]{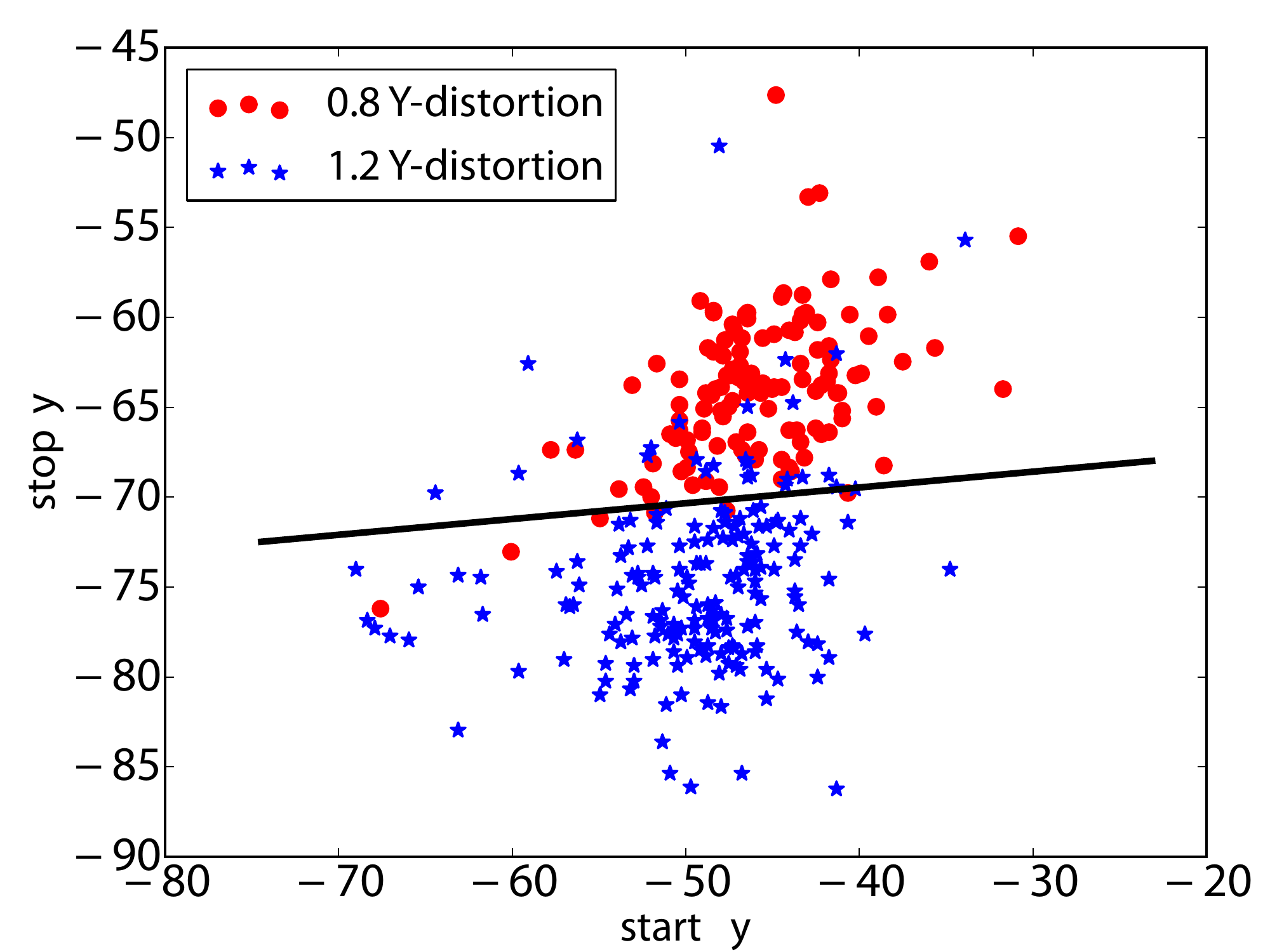}}
\subfloat[User C]{\includegraphics[width=0.33\textwidth]{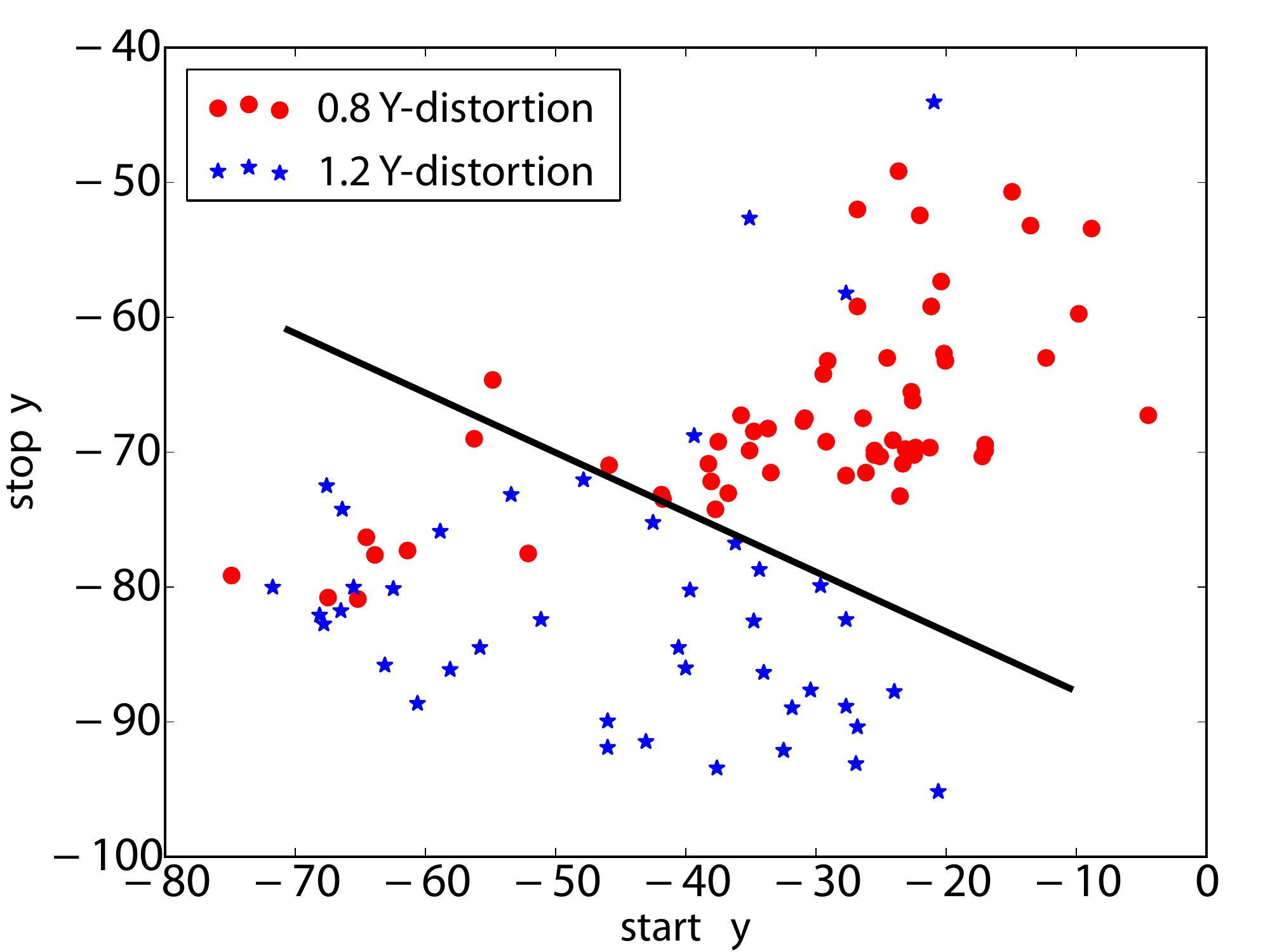}}
\caption{Start locations vs. stop locations in the vertical direction of  up strokes 
in  two screen settings of the three users. We find that their touch behaviors
are also \emph{sensitive}, i.e.,  a user's touch
strokes in different settings have a high degree of separability in the feature space.}
\label{sensitive} 
\end{figure*}

Before introducing our adaptive touch-based continuous authentication system, we
introduce two key observations that inspire the design of our authentication
system.

\myparatight{Stability and sensitivity} A user's touch behaviors in two screen
settings are said to be \emph{stable} if the touch behaviors of the user in one
setting are closer to those of the user in the other setting than those of other
users in both settings. Moreover, a user's touch behaviors in two settings are
said to be \emph{sensitive} if they have a high degree of separability in the
feature space. Figure~\ref{sta-sen} illustrates the possible outcomes for two
settings $s_a$ and $s_b$. Intuitively,  a user's touch behaviors in two settings become more
stable and less sensitive when the two settings are closer. 

We find that there exists screen settings across which a user's touch behaviors
are both stable and sensitive.   For instance, in our experiments (see
Section~\ref{sec:exp}), two of the screen settings we consider are 
0.8 Y-distortion and 1.2 Y-distortion,
respectively.  Figure~\ref{stroke} shows vertical touch strokes of three users
in the two settings; it is visually noticeable that their behaviors are stable. 

Note that a vertical stroke could be an \emph{up} stroke or a \emph{down} stroke, 
which corresponds to  scrolling up or scrolling down, respectively. 
Figure~\ref{sensitive} contrasts the start locations (i.e., start $y$)
and stop locations (i.e., stop $y$) in the vertical direction (i.e., Y axis)
of up strokes. We find that
their touch behavior are also sensitive. User A (or B) starts touch strokes at
similarly low $y$ locations in both settings. However, in the screen setting of 0.8
Y-distortion, the strokes interpreted by the application
are shorter than the strokes inputed by the user on the screen, and
thus the user automatically stops the touch strokes at higher (i.e., larger) $y$
locations, which makes the strokes on the screen longer. User C might subliminally notice the 
different screen settings, and
she tends to start and  stop the touch strokes at higher $y$ locations with 0.8 Y-distortion.

\myparatight{Implications} On one hand, stability implies that models trained in
any setting of one user will always be distinguishable from models trained for
other users, clearly separating individual users with high probabilities. This
property is needed to successfully authenticate a specific user. On the other
hand,  if we learn a model to distinguish a user's touch behaviors from other
users' in a single setting, then forgery attacks that replay the targeted user's
strokes collected in other settings can still succeed with high probabilities.

Sensitivity implies that we can train a model to distinguish a single user's touch behaviors
in one setting from those in  other settings, which makes it possible to defend against targeted attacks. 
Specifically, when our authentication system uses a setting $s$,  the attacker would fail to pass the authentication
with much higher probability if the attacker forges attacks using the targeted user's strokes collected in settings other than $s$.
\section{Adaptive Authentication}

Our touch-based authentication system consists of two phases: the
\emph{registration phase} and the \emph{authentication phase}. The
authentication system learns user characteristics during the registration phase
and enforces these learned characteristics during the authentication phase.

\begin{figure}[!t]
\centering
\includegraphics[width=0.45\textwidth]{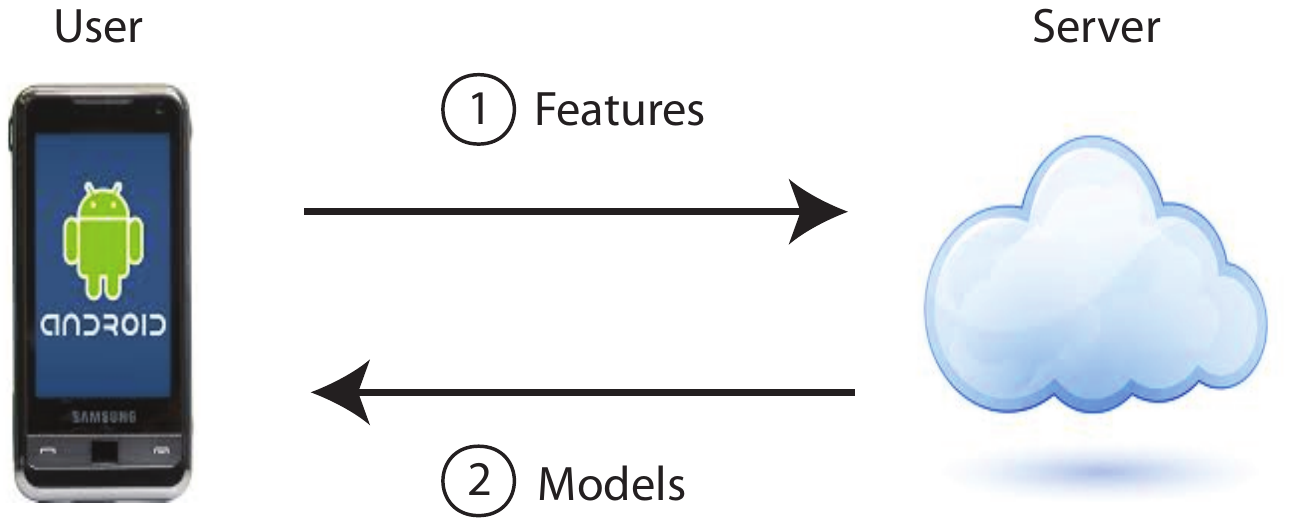}
\caption{Registration phase.  }
\label{registration}
\end{figure}


\subsection{Registration phase}

Figure~\ref{registration} illustrates the registration phase.  Suppose $u$ is a new user.  
First, we sample $n$ screen settings across which $u$'s touch behaviors are both stable and sensitive. In
 practice, we can evenly divide the range of possible screen settings into $m$ bins, and choose the centers of
 the $n$ bins that are randomly sampled.
 We denote the set of sampled settings as
$S(u)$.  

We distinguish two types of strokes: \emph{horizontal} and \emph{vertical}, 
which correspond to scrolling  horizontally and scrolling vertically, respectively. 
The two types of strokes have different features, e.g., the directions of the end-to-end lines corresponding to the strokes. This categorization could enable us to enhance the performance of the system.
 For each setting $s\in S(u)$, we collect a set of touch strokes for each type $t$
(denoted as $T(u, s, t)$) from $u$ via fixing the screen setting to be $s$. Then we
extract a set of features from each stroke. We adopt the features that are
proposed by Frank et al.~\cite{frank2013touchalytics}. The extracted features
are subsequently sent to the server.  We send features to the server instead of
the raw strokes to mitigate the loss from a server compromise. In particular, if we send
raw strokes to the server and it is compromised, the raw strokes are easily
available to attackers. This is similar to the scenario where passwords are
available to attackers when a password database is
compromised~\cite{rockyou, CSDN}. However, even if the server is compromised and
the features are available to attackers, it is unclear how to forge touch
strokes that have these features. 

Second, we leverage machine learning techniques to train a classifier for each
setting and each stroke type. Specifically, for each setting $s\in S(u)$ and a stroke type $t$, we take the features
of the corresponding strokes collected in the setting $s$ (i.e.,  $T(u, s, t)$) as positive
examples, and features of type-$t$ strokes collected in all other settings (i.e., 
$T(u,s', t)$ for all $s'\in S(u)-\{s\}$) and features of type-$t$ strokes of all users that have already adopted the system as
negative examples. Then we learn a classifier $c(u,s,t)$ to distinguish these
positive and negative examples. Therefore, we obtain $2n$ classifiers for
the user. The model parameters of these classifiers are then sent back to
the user. 

\subsection{Authentication phase}
Our adaptive continuous authentication method works on discrete time intervals. 
In each time interval,  we sample a setting $s$ from $S(u)$
uniformly at random and change the screen setting to be $s$. If the user inputs a stroke with type $t$, we  authenticate the stroke using the classifier $c(u, s, t)$.   


Intuitively, our authentication method can significantly decrease the success rates 
of 
targeted attacks.  In the worst case, the attacker obtains a set of touch
strokes of the user in \emph{all} settings in $S(u)$ and can replay these touch
strokes via programming a robot. However, the attacker cannot know which setting is randomly sampled in
a given time interval, and thus the attack is reduced to randomly guess a setting and 
program a robot to replay strokes collected in the setting. 
Due to the sensitivity of users' touch behaviors,  these forged  
strokes will be rejected with high probabilities.

Note that previous work~\cite{frank2013touchalytics,li2013unobservable} uses a fixed universal setting for
 all users, and thus their authentication systems can be breached if the
attacker obtains the targeted user's touch strokes in this hard-coded setting. Moreover, we
show (in Section~\ref{sec:exp}) that even replaying the touch strokes of
the targeted user collected in a different setting can still breach their
authentication system with high success rates because of the stability of users' touch behaviors.

A user's behavior is relatively stable  over time. For instance, Frank et
al.~\cite{frank2013touchalytics} showed that the median EER increases by only
4\% when their classifiers are used  one week after the registration phase.
However, in practice, to account for behavior variety in a longer period of
time,  we could periodically (e.g., each month or quarter) execute a
registration phase to update classifiers.  This is feasible since the
registration phase takes a short period of time as we will show in our
experiments.




\section{Experiments}
\label{sec:exp}

We evaluate the security of our new touch-based authentication system against forgery attacks 
and compare it with previous touch-based authentication systems.

\subsection{Data collection}

\begin{table}[!t]\renewcommand{\arraystretch}{1}
\centering
\caption{Notations of the five screen settings, which are distortions along either the X axis or the Y axis.}
\begin{tabular}{|c|c|c|c|c|} \hline
{\small $s_a$}&{\small $s_b$}&{\small $s_c$}&{\small $s_d$}&{\small $s_e$} \\ \hline
{\small 0.8}&{\small 0.9}&{\small 1.0}&{\small 1.1}&{\small 1.2} \\ \hline
\end{tabular}
\label{notation}
\end{table}

We consider five screen settings: 0.8, 0.9, 1.0, 1.1, and 1.2 distortions along either
the X or Y axis, and we denote them as $s_a$, $s_b$, $s_c$, $s_d$, and $s_e$, respectively. 
Table~\ref{notation} shows the notations of the five screen settings. 
We choose these settings because they  are reasonably separable from each other so that 
a user's touch behaviors are sensitive, yet transitions between them are still
unnoticeable to users.  

We aim to collect horizontal strokes and vertical strokes in the five screen settings.
Moreover, we want to study if transitions between screen settings can be 
performed without users noticing them.   
To achieve these goals, we implemented an Android application with two games to 
record users' raw touch data.  
Our application uses an Android API  to configure screen settings.
Moreover, we implemented a library to intercept the raw touch data. 
Considering user fatigue, the ordering of the two games  is shuffled uniformly at random.

\begin{figure}[!t]
\centering
\subfloat[Task 1]{\includegraphics[width=0.45\columnwidth]{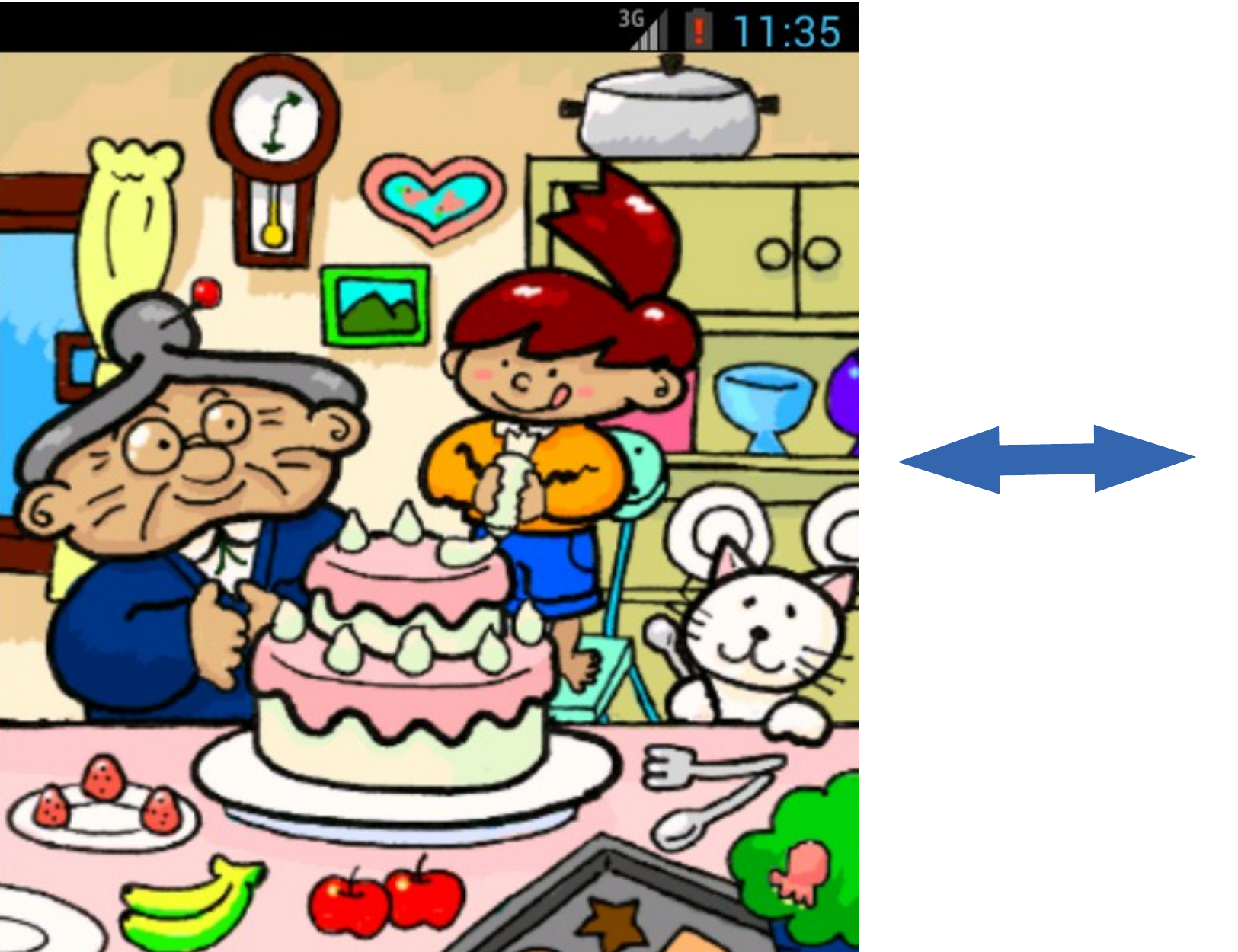}
\label{fig:horizontal-game}}
\subfloat[Task 2]{\includegraphics[width=0.45\columnwidth]{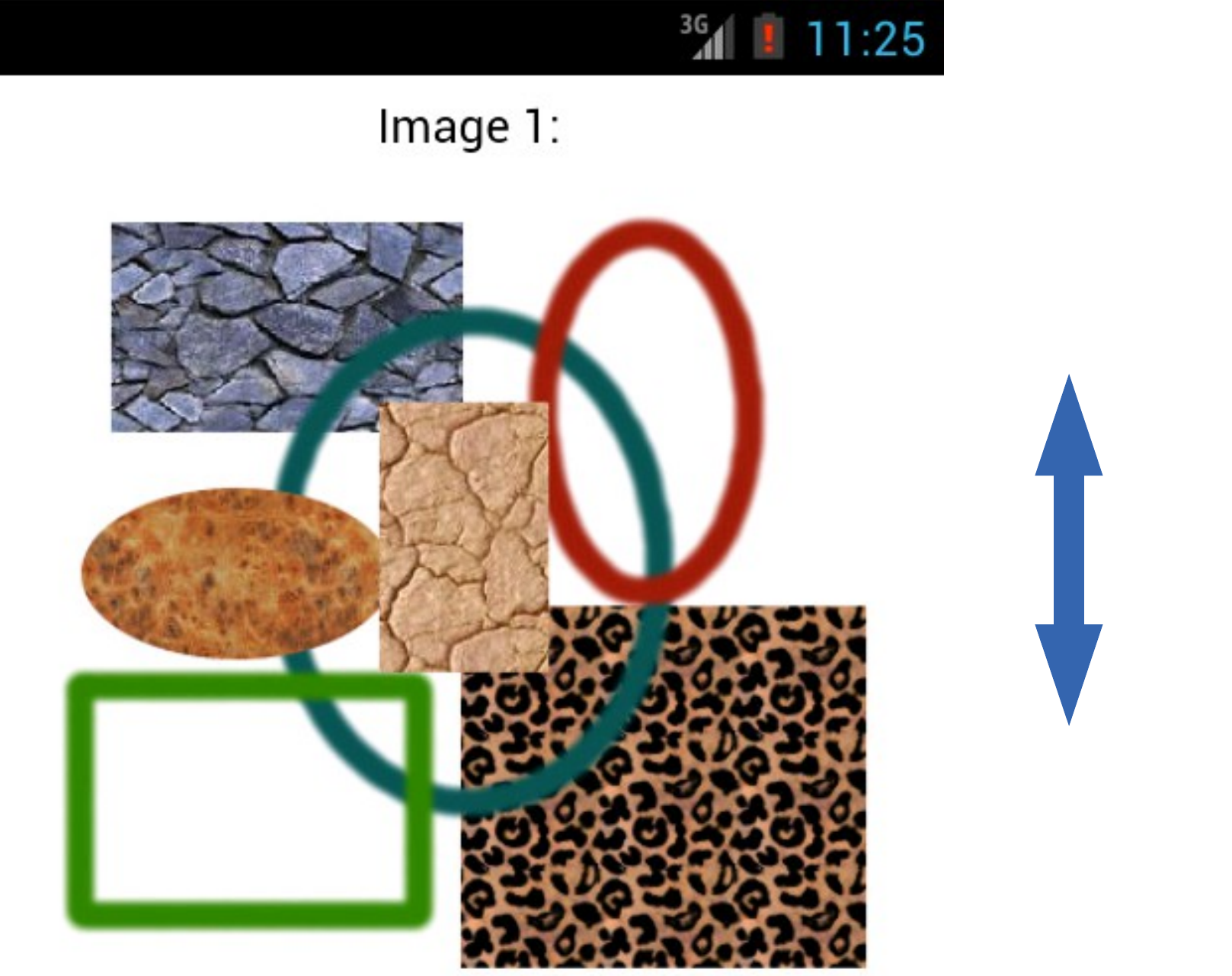}
\label{fig:vertical-game}}
\caption{(a) The game used in the task 1; the user must swipe
horizontally. The image itself is available under the Creative Commons
Attribution-Share Alike 3.0 license from the Wikipedia website. 
(b) The game used in the task 2; the user must swipe vertically.
We produced this image and release it under the Creative Commons Attribution
Share-Alike 3.0 license.}
\vspace{6mm}
\end{figure}

\myparatight{Task 1, horizontal strokes} In the first game the user must identify differences between two images. The
application shows two versions of one image in a horizontal gallery with a black
image in between. The user must swipe  horizontally between the images. 
Figure~\ref{fig:horizontal-game} shows a screenshot of this game. 
We shuffle the five screen settings along the X axis at the beginning of the game, and change a setting every 30s. 

 \begin{table}[!t]\renewcommand{\arraystretch}{1}
\centering
\caption{ The number of strokes per subject and per second in the five screen settings for the 25 subjects.}
\addtolength{\tabcolsep}{-2pt}
\begin{tabular}{|c|c|c|c|c|c|c|c|c|c|c|} \hline
& \multicolumn{5}{|c|}{{\small Strokes per subject}} & \multicolumn{5}{|c|}{{\small Strokes per second}} \\ \cline{2-11}
&{\small $s_a$}&{\small $s_b$}&{\small $s_c$}&{\small $s_d$}&{\small $s_e$}&{\small $s_a$}&{\small $s_b$}&{\small $s_c$}&{\small $s_d$}&{\small $s_e$} \\ \hline
{\small Horizontal}&{\small 51}&{\small 52}&{\small 48}&{\small 49}&{\small 52}&{\small 0.35}&{\small 0.30}&{\small 0.19}&{\small 0.17}&{\small 0.17}\\ \hline
{\small Vertical}&{\small 83}&{\small 77}&{\small 74}&{\small 78}&{\small 75}&{\small 0.71}&{\small 0.47}&{\small 0.31}&{\small 0.30}&{\small 0.27}\\ \hline
\end{tabular}
\label{dataset}
\vspace{3mm}
\end{table}

\myparatight{Task 2, vertical strokes}
In the second game the user must identify pairs of images. The application shows
five (statically pre-shuffled) pairs of vertically aligned images of different
patterns, shapes, and colors. At one given time only one image is visible and
the user must move the screen to other images along the Y axis. The static shuffling
allows us to compare the same task across different users.
Figure~\ref{fig:vertical-game}  shows a screenshot of this game. 
We shuffle the five screen settings along the Y axis at the beginning of the game, 
and change a setting every 30s.

We used a HTC One smartphone to collect data from 25 subjects. 
These subjects age between 25 and 35; and 11 of them are female.  
Table~\ref{dataset} summarizes our
dataset.

\myparatight{Smooth transition between screen settings}
Our authentication system automatically transits from one
setting to another in different time intervals. One natural question is that if
users notice such transitions. To answer this question, we ask each subject
the question ``Did you notice anything abnormal?" 
when he/she  finishes the tasks.

We find that \emph{no} subject noticed these transitions, which means that 
users  subconsciously adapt their behavior to different screen settings and 
transitions between settings do not interrupt users nor influence user
experiences.

\begin{table*}[!t]\renewcommand{\arraystretch}{1.1}
\centering
\caption{Approaches we compare.}
\addtolength{\tabcolsep}{-5pt}
\begin{tabular}{|c|c|} \hline
{\small Notation} & {\small Description} \\ \hline
{\small C-Baseline-$x$} & {\small Baseline classifier only using touch strokes in the setting $s_x$, where $x\in \{a,b,c,d,e\}$} \\ \hline
{\small C-ATCA-$x$} & {\small Our classifier for the setting $s_x$, where $x\in \{a,b,c,d,e\}$; strokes in other settings are also used} \\ \hline
{\small S-Baseline-$x$} & {\small Baseline authentication system using C-Baseline-$x$ with the fixed universal setting $s_x$, where $x\in \{a,b,c,d,e\}$ }\\ \hline
{\small S-Baseline-improved} & {\small Our improved Baseline authentication system whose setting is randomly selected} \\ \hline
\multirow{2}{*}{{\small S-ATCA}} &{\small  Our authentication system using C-ATCA-$x$, where $x\in \{a,b,c,d,e\}$;}\\ 
&{\small  settings are randomly selected in each time interval} \\ \hline
\end{tabular}
\label{approach}
\end{table*}

\subsection{Experimental setups} 
We evaluate both random attacks (RA) and targeted attacks (TA) against previous
approaches~\cite{frank2013touchalytics,li2013unobservable,attack-CCS13} and
ours; we explore the impact of the number of screen settings on the performance of our approach; 
and  we study the time required to collect training strokes in the registration phase.


\subsubsection{Compared approaches} 
We distinguish classifiers and authentication systems. Classifiers are key
components of a touch-based authentication system, but an authentication system
also requires to consider screen settings. We name  approaches to train
classifiers with a prefix 'C' and approaches to implement authentication systems
with a prefix 'S'. Table~\ref{approach} summarizes the approaches we compare.

\myparatight{Classifiers} We compare the following classifiers:
\begin{packeditemize}
\item {\bf
C-Baseline~\cite{frank2013touchalytics,li2013unobservable,attack-CCS13}:} This
approach considers one setting.  To train a classifier for a user $u$, they take
the touch strokes of $u$ in a setting as positive examples and those of all other
users in the same setting as negative examples. We use C-Baseline-$x$ to denote 
their approach in the setting $s_x$, where $x\in\{a, b, c,d,e\}$.

\item {\bf C-ATCA:} Our adaptive touch-based continuous authentication (ATCA)
considers multiple settings when training classifiers. Specifically, for a user
$u$, we train  classifiers for all the five screen settings. We denote them as C-ATCA-a, 
C-ATCA-b, C-ATCA-c, C-ATCA-d, and C-ATCA-e, respectively. 
To train C-ATCA-$x$, we take  strokes of
$u$ in the setting $s_x$  as positive examples, and strokes of $u$ in
the other four settings and strokes of other users in all the five settings
as negative examples, where $x\in\{a, b, c,d,e\}$.
\end{packeditemize}

We adopt Support Vector Machine (SVM)~\cite{svm2} as the classifier~\cite{svm2}
in all compared approaches since it was shown to perform well by previous
work~\cite{frank2013touchalytics,li2013unobservable}. 

\myparatight{Authentication systems} Other than using different classifiers,
authentication systems might also have different ways to use screen settings.

\begin{packeditemize}

\item {\bf S-Baseline~\cite{frank2013touchalytics, li2013unobservable,
attack-CCS13}:}  Their approach uses an universal setting (e.g., the default
setting of the smartphone system) for all users. In this case, the attacker can
obtain the universal setting, e.g., via reading the code of the authentication
system. We further use \emph{S-Baseline-$x$} to denote the
authentication system with the universal setting $s_x$, where $x\in\{a, b, c,d,e\}$.
Note that S-Baseline-$x$ uses the classifier C-Baseline-$x$.  

\item {\bf S-Baseline-improved:} We improve S-Baseline via selecting a setting
from $\{s_a,s_b,s_c,s_d,s_e\}$ uniformly at random in the registration phase and fixing it
in the authentication phase for each user. The S-Baseline-improved
 system makes the attacker unaware of the setting used for a
targeted user. 

\item {\bf S-ATCA:} Our adaptive touch-based continuous authentication (ATCA)
selects a setting $s_x$ from the considered five settings uniformly at random in each time
interval and uses the classifier  C-ATCA-$x$ to authenticate users, where
$x\in\{a,b,c,d,e\}$.  

\end{packeditemize}



\begin{table*}[!t]\renewcommand{\arraystretch}{1}
\centering
\caption{Mean EERs over all subjects for each classifier and attack dataset for
horizontal strokes. Numbers in parentheses are standard deviations.}
\subfloat[Horizontal strokes, random attacks]{
\begin{tabular}{|c|c|c|c|c|c|} \hline
&{\small $s_a$} & {\small $s_b$} & {\small $s_c$} & {\small $s_d$} & {\small $s_e$} \\ \hline
{\small C-Baseline-a}&{\small 0.06(0.0543)}&{\small 0.06(0.0485)}&{\small 0.06(0.0472)}&{\small 0.07(0.0530)}&{\small 0.07(0.0547)}\\ \hline
{\small C-Baseline-b}&{\small 0.04(0.0461)}&{\small 0.04(0.0462)}&{\small 0.05(0.0423)}&{\small 0.05(0.0447)}&{\small 0.05(0.0472)}\\ \hline
{\small C-Baseline-c}&{\small 0.07(0.0590)}&{\small 0.07(0.0518)}&{\small 0.06(0.0498)}&{\small 0.06(0.0463)}&{\small 0.06(0.0541)}\\ \hline
{\small C-Baseline-d}&{\small 0.10(0.0817)}&{\small 0.09(0.0818)}&{\small 0.09(0.0742)}&{\small 0.09(0.0792)}&{\small 0.09(0.0757)}\\ \hline
{\small C-Baseline-e}&{\small 0.09(0.0891)}&{\small 0.10(0.0966)}&{\small 0.10(0.0942)}&{\small 0.10(0.0973)}&{\small 0.09(0.0991)}\\ \hline
{\small C-ATCA-a}&{\small 0.04(0.0736)}&{\small \bf 0.02(0.0534)}&{\small \bf 0.02(0.0437)}&{\small \bf 0.01(0.0355)}&{\small \bf 0.01(0.0291)}\\ \hline
{\small C-ATCA-b}&{\small \bf 0.02(0.0554)}&{\small 0.04(0.0681)}&{\small \bf 0.02(0.0572)}&{\small 0.02(0.0437)}&{\small 0.02(0.0544)}\\ \hline
{\small C-ATCA-c}&{\small 0.06(0.0782)}&{\small 0.07(0.0822)}&{\small 0.08(0.0778)}&{\small 0.06(0.0634)}&{\small 0.06(0.0675)}\\ \hline
{\small C-ATCA-d}&{\small 0.06(0.0771)}&{\small 0.06(0.0823)}&{\small 0.06(0.0787)}&{\small 0.09(0.0857)}&{\small 0.07(0.0811)}\\ \hline
{\small C-ATCA-e}&{\small 0.03(0.0719)}&{\small 0.04(0.0743)}&{\small 0.04(0.0703)}&{\small 0.05(0.0755)}&{\small 0.06(0.0864)}\\ \hline
\end{tabular}
\label{hor-ra}
}

\subfloat[Horizontal strokes, targeted attacks]{
\begin{tabular}{|c|c|c|c|c|c|} \hline
&{\small $s_a$} & {\small $s_b$} & {\small $s_c$} & {\small $s_d$} & {\small $s_e$} \\ \hline
{\small C-Baseline-a}&{\small 0.50(0.0000)}&{\small 0.49(0.1142)}&{\small 0.44(0.1586)}&{\small 0.38(0.2023)}&{\small 0.37(0.2051)}\\ \hline
{\small C-Baseline-b}&{\small 0.46(0.1142)}&{\small 0.50(0.0000)}&{\small 0.47(0.1388)}&{\small 0.35(0.1602)}&{\small 0.35(0.1710)}\\ \hline
{\small C-Baseline-c}&{\small 0.49(0.1442)}&{\small 0.49(0.1181)}&{\small 0.50(0.0000)}&{\small 0.46(0.1397)}&{\small 0.41(0.1876)}\\ \hline
{\small C-Baseline-d}&{\small 0.43(0.1926)}&{\small 0.44(0.1866)}&{\small 0.46(0.1376)}&{\small 0.50(0.0000)}&{\small 0.47(0.1129)}\\ \hline
{\small C-Baseline-e}&{\small 0.39(0.2480)}&{\small 0.40(0.2210)}&{\small 0.41(0.1852)}&{\small 0.42(0.1752)}&{\small 0.50(0.0000)}\\ \hline
{\small C-ATCA-a}&{\small 0.50(0.0000)}&{\small 0.27(0.1563)}&{\small \bf 0.23(0.1384)}&{\small \bf 0.16(0.1684)}&{\small \bf 0.16(0.1515)}\\ \hline
{\small C-ATCA-b}&{\small 0.33(0.1475)}&{\small 0.50(0.0000)}&{\small 0.30(0.1166)}&{\small 0.25(0.1185)}&{\small 0.23(0.1437)}\\ \hline
{\small C-ATCA-c}&{\small 0.37(0.1569)}&{\small 0.35(0.1340)}&{\small 0.50(0.0000)}&{\small 0.32(0.1100)}&{\small 0.35(0.1370)}\\ \hline
{\small C-ATCA-d}&{\small 0.27(0.1390)}&{\small 0.29(0.1353)}&{\small 0.33(0.1311)}&{\small 0.50(0.0000)}&{\small 0.35(0.1019)}\\ \hline
{\small C-ATCA-e}&{\small \bf 0.21(0.1780)}&{\small \bf 0.22(0.1729)}&{\small 0.24(0.1721)}&{\small 0.25(0.1516)}&{\small 0.50(0.0000)}\\ \hline
\end{tabular}
\label{hor-ta}
}
\label{hor}
\vspace{-4mm}
\end{table*}

\subsubsection{Training and testing} 
We evaluate the approaches via 5-fold cross-validation.
  Next, we take \emph{horizontal strokes} as an example to
illustrate the details. The vertical strokes are treated in the same way. The set of
our subjects is denoted as $U_d$.

We evenly split horizontal strokes of each user in each setting into 5
folds uniformly at random. Let $F=\{1,2,3,4,5\}$ denote the IDs of the 5 folds.
Moreover, we denote by $f(u,s,i)$ the $i$th fold of horizontal strokes of the user $u$ in
the setting $s$, where $i\in F$ and $s\in \{s_a,s_b,s_c,s_d,s_e\}$. 

For each user $u$, we iterate over $i$. For each $i$, we train classifiers as
follows:

\myparatight{Training C-Baseline classifiers} To train C-Baseline-$x$, we
use $u$'s horizontal strokes in the setting $s_x$ (i.e., $\cup_{j\in
F-\{i\}}$ $f(u,s_x,j)$ as positive examples and other users' horizontal strokes in the
setting $s_x$ (i.e., $\cup_{v\in U_d-\{u\}}\cup_{j\in F-\{i\}}f(u,s_x,j)$) as negative examples, where $x\in\{a,b,c,d,e\}$.  

\myparatight{Training C-ATCA classifiers} To train a C-ATCA-$x$ classifier,
we use $u$'s horizontal strokes in the setting $s_x$ (i.e., $\cup_{j\in
F-\{i\}}f(u,s_x,j)$) as positive examples. However, we treat $u$'s horizontal strokes
in the other four settings (i.e., $\cup_{j\in F-\{i\}}f(u,s_y,j)$, where $y\in \{a,b,c,d,e\}/$
$\{x\}$) and other users'
horizontal strokes in all the five settings (i.e., $\cup_{v\in U_d-\{u\}}\cup_{j\in
F-\{i\}}f(u,s_z,j)$, where $z\in\{a,b,c,d,e\}$)
as negative examples, where $x\in\{a,b,c,$ 
$d,e\}$. 

We adopt a Gaussian kernel for SVM and use the LibSVM
library~\cite{chang2011libsvm} to learn the corresponding hyper-parameters via
grid search. Each feature is re-scaled to be between -1 and 1. Note that
training features and testing features are normalized separately.

\myparatight{Testing} In the test phase, we use $u$'s horizontal strokes in a setting
(i.e., $f(u,s_x,i)$) as legitimate (or positive) examples for the classifiers
trained for the same setting (i.e., C-Baseline-$x$ and C-ATCA-$x$), where $x\in
\{a,b,c,d,e\}$. Moreover, we treat other users' horizontal strokes in all the five settings (i.e.,
$\cup_{v\in U_d-\{u\}} f(v,s_z,i)$, where $z\in\{a,b,c,d,e\}$) as
strokes to perform random attacks; and we treat the target user's horizontal strokes
in the five settings (i.e., $f(u,s_z,i)$, where $z\in\{a,b,c,d,$ 
$e\}$) as strokes to perform
targeted attacks. For each user, training and testing are performed with 5 \emph{trials} since
we have 5 folds, and the results are averaged over them.

\subsubsection{Evaluation metrics}
Our evaluation metric involves the false-acceptance rate (FAR), the false
rejection rate (FRR), and the mean time $T$ required to make the first
authentication decision in a session. FAR is the fraction of strokes of
imposters that are recognized as strokes of the legitimate user by the classifier.
FRR is the fraction of strokes of legitimate users that are rejected by the
classifier. FRR quantifies the empirical probability that the legitimate user must
resort to conventional authentication mechanisms. Put in a temporal context, if
$T_s$ is the average time between two strokes, then the expected time after
which the legitimate user must type in a password due to misclassification is
FRR$^{-1}T_s$.

The two error rates FRR and FAR can be traded off against each other via 
changing the decision threshold of the classifier. For instance, at the
cost of missing out some imposters one can reduce FRR by decreasing the threshold.
In order to account for this
usability-security trade-off, we report the equal error rate (EER) in all
experiments. This is the error rate at the threshold where
FAR equals FRR.

For a classifier and an attack dataset (e.g., random attacks using strokes
collected in $s_a$,  targeted attacks using strokes collected in $s_b$), we compute an EER using the dataset that consists of the
classifier's test positive examples (legitimate  strokes) and the attack
dataset. Intuitively, EER in our context measures the degree of separability
between  touch strokes of legitimate users and attack strokes.

 

\subsection{Results for classifiers}

Table~\ref{hor} and Table~\ref{ver} show the mean EERs of our subjects for each classifier and
each attack dataset for horizontal strokes and vertical strokes, respectively.



\myparatight{Diff-setting attacks vs. same-setting attacks} We call an attack as
a \emph{same-setting attack} if the attack strokes are collected in the same
setting with the one in which the classifier uses. Otherwise, we call an attack
\emph{diff-setting attack}. For instance, random attacks using strokes collected
in the setting $s_a$ to the classifier C-Baseline-a or C-ATCA-a are same-setting
attacks, while random attacks using strokes collected in the setting $s_a$ to
the classifier C-Baseline-b or C-ATCA-b are diff-setting attacks. Moreover, 
we denote by RA-$xy$ (or TA-$xy$) the random attacks (or targeted attacks) that use
strokes collected in the setting $s_y$ to the classifier that uses
the setting $s_x$, where $x,y\in\{a,b,c,d,e\}$.

\begin{table*}[!t]\renewcommand{\arraystretch}{1}
\centering
\caption{Mean EERs over all subjects for each classifier and attack dataset for
vertical strokes. Numbers in parentheses are standard deviations.}
\subfloat[Vertical strokes, random attacks]{
\begin{tabular}{|c|c|c|c|c|c|} \hline
&{\small $s_a$} & {\small $s_b$} & {\small $s_c$} & {\small $s_d$} & {\small $s_e$} \\ \hline
{\small C-Baseline-a}&{\small 0.09(0.0873)}&{\small 0.10(0.0913)}&{\small 0.11(0.0986)}&{\small 0.11(0.1024)}&{\small 0.12(0.1089)}\\ \hline
{\small C-Baseline-b}&{\small 0.08(0.0641)}&{\small 0.08(0.0652)}&{\small 0.09(0.0697)}&{\small 0.10(0.0760)}&{\small 0.10(0.0867)}\\ \hline
{\small C-Baseline-c}&{\small 0.11(0.0992)}&{\small 0.12(0.1019)}&{\small 0.12(0.1032)}&{\small 0.12(0.1091)}&{\small 0.13(0.1161)}\\ \hline
{\small C-Baseline-d}&{\small 0.12(0.1019)}&{\small 0.12(0.0969)}&{\small 0.12(0.0980)}&{\small 0.11(0.0936)}&{\small 0.12(0.1022)}\\ \hline
{\small C-Baseline-e}&{\small 0.15(0.1100)}&{\small 0.14(0.1046)}&{\small 0.14(0.1065)}&{\small 0.14(0.1060)}&{\small 0.15(0.1158)}\\ \hline
{\small C-ATCA-a}&{\small \bf 0.07(0.0802)}&{\small \bf 0.05(0.0727)}&{\small \bf 0.05(0.0697)}&{\small \bf 0.05(0.0692)}&{\small \bf 0.07(0.0960)}\\ \hline
{\small C-ATCA-b}&{\small 0.08(0.0698)}&{\small 0.11(0.0742)}&{\small 0.09(0.0742)}&{\small 0.08(0.0766)}&{\small 0.09(0.0819)}\\ \hline
{\small C-ATCA-c}&{\small 0.08(0.0748)}&{\small 0.08(0.0785)}&{\small 0.12(0.0914)}&{\small 0.09(0.0799)}&{\small 0.11(0.0899)}\\ \hline
{\small C-ATCA-d}&{\small \bf 0.07(0.0688)}&{\small 0.08(0.0763)}&{\small 0.08(0.0759)}&{\small 0.11(0.0843)}&{\small 0.10(0.0846)}\\ \hline
{\small C-ATCA-e}&{\small \bf 0.07(0.0702)}&{\small 0.06(0.0644)}&{\small 0.07(0.0672)}&{\small 0.08(0.0728)}&{\small 0.12(0.0935)}\\ \hline
\end{tabular}
\label{ver-ra}
}

\subfloat[Vertical strokes, targeted attacks]{
\begin{tabular}{|c|c|c|c|c|c|} \hline
&{\small $s_a$} & {\small $s_b$} & {\small $s_c$} & {\small $s_d$} & {\small $s_e$} \\ \hline
{\small C-Baseline-a}&{\small 0.50(0.0000)}&{\small 0.44(0.0930)}&{\small 0.42(0.1322)}&{\small 0.41(0.1587)}&{\small 0.35(0.1651)}\\ \hline
{\small C-Baseline-b}&{\small 0.48(0.0747)}&{\small 0.50(0.0000)}&{\small 0.45(0.1198)}&{\small 0.42(0.1209)}&{\small 0.38(0.1289)}\\ \hline
{\small C-Baseline-c}&{\small 0.44(0.1189)}&{\small 0.44(0.0825)}&{\small 0.50(0.0000)}&{\small 0.46(0.1199)}&{\small 0.41(0.1614)}\\ \hline
{\small C-Baseline-d}&{\small 0.42(0.1345)}&{\small 0.43(0.0978)}&{\small 0.46(0.1154)}&{\small 0.50(0.0000)}&{\small 0.44(0.0811)}\\ \hline
{\small C-Baseline-e}&{\small 0.44(0.1018)}&{\small 0.44(0.1122)}&{\small 0.45(0.0820)}&{\small 0.47(0.1025)}&{\small 0.50(0.0000)}\\ \hline
{\small C-ATCA-a}&{\small 0.50(0.0000)}&{\small \bf 0.26(0.0956)}&{\small \bf 0.26(0.1264)}&{\small \bf 0.24(0.1304)}&{\small \bf 0.23(0.1405)}\\ \hline
{\small C-ATCA-b}&{\small 0.36(0.1200)}&{\small 0.50(0.0000)}&{\small 0.30(0.0940)}&{\small 0.31(0.1167)}&{\small 0.31(0.1324)}\\ \hline
{\small C-ATCA-c}&{\small 0.33(0.1093)}&{\small 0.32(0.1048)}&{\small 0.50(0.0000)}&{\small 0.31(0.1091)}&{\small 0.32(0.1282)}\\ \hline
{\small C-ATCA-d}&{\small 0.28(0.0949)}&{\small 0.31(0.0842)}&{\small 0.30(0.0986)}&{\small 0.50(0.0000)}&{\small 0.31(0.0803)}\\ \hline
{\small C-ATCA-e}&{\small \bf 0.25(0.1091)}&{\small 0.28(0.0988)}&{\small 0.30(0.0811)}&{\small 0.29(0.0914)}&{\small 0.50(0.0000)}\\ \hline
\end{tabular}
\label{ver-ta}
}
\label{ver}

\end{table*}

\begin{table*}[!t]\renewcommand{\arraystretch}{1}
\centering
\caption{Possible attacks to the 7  authentication systems.}
\addtolength{\tabcolsep}{-3pt}
\begin{tabular}{|c|c|c|} \hline
 & {\small Random attacks} & {\small Targeted attacks} \\ \hline
{\small S-Baseline-$a$} & {\small $\max\{\text{RA-}ay\}$ for $y\in \{a,b,c,d,e\}$} & {\small TA-$aa$} \\ \hline
{\small S-Baseline-$b$} & {\small $\max\{\text{RA-}by\}$ for $y\in \{a,b,c,d,e\}$} & {\small TA-$bb$} \\ \hline
{\small S-Baseline-$c$} & {\small $\max\{\text{RA-}cy\}$ for $y\in \{a,b,c,d,e\}$} & {\small TA-$cc$} \\ \hline
{\small S-Baseline-$d$} & {\small $\max\{\text{RA-}dy\}$ for $y\in \{a,b,c,d,e\}$} & {\small TA-$dd$} \\ \hline
{\small S-Baseline-$e$} & {\small $\max\{\text{RA-}ey\}$ for $y\in \{a,b,c,d,e\}$} & {\small TA-$ee$} \\ \hline
{\small S-Baseline-improved} & {\small RA-$xy$, where $x,y\in \{a,b,c,d,e\}$} & {\small TA-$xy$, where $x,y\in \{a,b,c,d,e\}$} \\ \hline
{\small S-ATCA} & {\small RA-$xy$, where $x,y\in \{a,b,c,d,e\}$} & {\small TA-$xy$, where $x,y\in \{a,b,c,d,e\}$} \\ \hline
\end{tabular}
\label{attacks}
\end{table*}

As we expect, same-setting targeted attacks achieve
higher EERs than diff-setting targeted attacks for both C-Baseline classifiers
 and our C-ATCA classifiers. This is because  users' touch behaviors are sensitive.
 For instance, for horizontal strokes,
 EERs of diff-setting targeted attacks are 13\%-34\% smaller than those of  
same-setting targeted attacks for our C-ATCA classifiers depending on which setting
is used to collect the targeted attacks data. 

Moreover, when the difference between the screen setting used to collect 
the targeted attacks data and the screen setting of the classifier increases, 
the EER of the corresponding diff-setting attacks decreases. For instance, for horizontal
strokes,
 the EER of the diff-setting targeted attack TA-$ea$ is 12\% smaller than that of 
the diff-setting targeted attack TA-$ba$ for our C-ATCA classifiers. Our observations imply that users'
touch behaviors are more sensitive when the differences between screen settings are larger.

\myparatight{C-Baseline vs. C-ATCA} Our classifiers
perform significantly better than C-Baseline classifiers at  defending
against diff-setting attacks. Specifically, EERs of diff-setting random attacks
to our classifiers are 1\% to 8\% smaller than those of the C-Baseline
classifiers. For instance, with horizontal strokes,  the EER of
the diff-setting random attacks RA-$ea$ is 9\% for the C-Baseline-$e$ classifier.
 However, the EER of RA-$ea$ is 3\% for 
our classifier C-ATCA-$e$, which is 6\% smaller than the C-Baseline-$e$ classifier.
 Moreover, EERs of diff-setting
targeted attacks to  our classifiers are 6\% to 22\% smaller than those of the
C-Baseline classifiers. For instance, with horizontal strokes, the
EER of the diff-setting targeted attacks TA-$ea$ is 39\% for the C-Baseline-$e$ classifier.
However, the EER of TA-$ea$ is 21\% for our classifier C-ATCA-$e$, which is 18\% smaller than the 
C-Baseline-$e$ classifier.
This is because a user's
touch behaviors in the five screen settings are both stable and sensitive, which results
in high EERs for the C-Baseline classifiers and explains the low EERs for our
classifiers, respectively.

For same-setting random attacks, the EERs of our classifiers are slightly larger
than those of the C-Baseline classifiers in some cases. This is because our classifiers in a
setting $s$ use more negative examples other than the strokes of other users
collected in $s$, which somehow makes their decision boundaries move towards the
strokes of other users collected in $s$, and thus same-setting random attacks
achieve slightly higher EERs. As we expect, same-setting targeted attacks
achieve high EERs for all classifiers. Specifically, EERs of our classifiers and
the C-Baseline classifiers are all close to 50\% for same-setting targeted
attacks. This means that, for each stroke, the classifier makes a
random decision, i.e., it accepts or rejects it with the same probability of
0.5.



\subsection{Results for authentication systems}
We first introduce possible attacks to the considered authentication systems and then show comparison results.

\myparatight{Attacks} Suppose 
the attacker already knows the set of
settings $\{s_a,s_b, s_c,s_d,s_e\}$ that \emph{could} be used by the authentication systems.
Moreover, for a targeted user, we assume the attacker obtains touch strokes of
the targeted user or other users in all the five settings. This means
that the attacker can perform targeted attacks or random attacks using strokes
collected in any of the five settings. 
 Recall that RA-$xy$ (or TA-$xy$) denotes the random attacks (or targeted attacks) that use
strokes collected in the setting $s_y$ to authentication systems that use
the setting  $s_x$, where $x,y\in\{a,b,c,d,e\}$.  
Note that the attacker does not know which setting (or classifier) is
currently used by our authentication system at a given time point.

%
%
%

To attack the baseline authentication system that uses the setting $s_x$ (i.e.,
S-Baseline-$x$), the attacker can choose to perform the best random attacks
(i.e., $\max\{\text{RA}$
$\text{-}xy\}$ for $y\in\{a,b,c,d,e\}$) or 
same-setting targeted attacks (i.e., TA-$xx$) that achieve the highest EERs,
 where
$x\in\{a,b,c,d,e\}$. This is because the attacker can know the used screen setting.


To attack S-Baseline-improved or S-ATCA, the attacker does not know the setting
of the authentication system\footnote{Note that we assume the attacker cannot
access the settings of the authentication system  at runtime, because such
access requires high privileges (e.g., root access to the operating system) and
an attacker that has already obtained these high privileges already compromised
the device.} and thus it randomly selects a setting and replays
strokes collected in the selected setting. As a result, the 25 possible
attacks RA-$xy$ for $x,y\in \{a,b,c,d,e\}$ are performed with an equal probability
of $\frac{1}{25}$ in the random attacks, and TA-$xy$ for $x,y\in \{a,b,c,d,e\}$ are
performed with an equal probability of $\frac{1}{25}$ in the targeted attacks. 
Table~\ref{attacks} summarizes the possible attacks to different authentication
systems.




\begin{table}[!t]\renewcommand{\arraystretch}{1}
\centering
\caption{Mean EERs over all subjects for each authentication system and attack
for horizontal strokes and vertical strokes. Numbers in parentheses are standard deviations. 
We find that our authentication system achieves 
significantly smaller EERs than previous work for both random attacks and targeted attacks. }
\addtolength{\tabcolsep}{-2pt}

\subfloat[Horizontal strokes]{
\begin{tabular}{|c|c|c|} \hline
& {\small Random attacks} & {\small Targeted attacks} \\ \hline
{\small S-Baseline-a}&{\small 0.08(0.0577)}&{\small 0.50(0.0000)}\\ \hline
{\small S-Baseline-b}&{\small 0.06(0.0516)}&{\small 0.50(0.0000)}\\ \hline
{\small S-Baseline-c}&{\small 0.09(0.0616)}&{\small 0.50(0.0000)}\\ \hline
{\small S-Baseline-d}&{\small 0.11(0.0817)}&{\small 0.50(0.0000)}\\ \hline
{\small S-Baseline-e}&{\small 0.11(0.0969)}&{\small 0.50(0.0000)}\\ \hline
{\small S-Baseline-improved}&{\small 0.07(0.0412)}&{\small 0.44(0.0512)}\\ \hline
{\small S-ATCA}&{\small \bf 0.04(0.0488)}&{\small \bf 0.32(0.0783)}\\ \hline
\end{tabular}
\label{horizontal-auth}}

\subfloat[Vertical strokes]{
\addtolength{\tabcolsep}{-0.5pt}
\begin{tabular}{|c|c|c|} \hline
& {\small Random attacks} & {\small Targeted attacks} \\ \hline
{\small S-Baseline-a}&{\small 0.12(0.1067)}&{\small 0.50(0.0000)}\\ \hline
{\small S-Baseline-b}&{\small 0.11(0.0819)}&{\small 0.50(0.0000)}\\ \hline
{\small S-Baseline-c}&{\small 0.14(0.1111)}&{\small 0.50(0.0000)}\\ \hline
{\small S-Baseline-d}&{\small 0.14(0.1051)}&{\small 0.50(0.0000)}\\ \hline
{\small S-Baseline-e}&{\small 0.17(0.1187)}&{\small 0.50(0.0000)}\\ \hline
{\small S-Baseline-improved}&{\small 0.12(0.0777)}&{\small 0.45(0.0364)}\\ \hline
{\small S-ATCA}&{\small \bf 0.08(0.0542)}&{\small \bf 0.33(0.0502)}\\ \hline
\end{tabular}
\label{vertical-auth}}
\label{result-auth}
\vspace{-4mm}
\end{table}

\begin{figure}[ht]
\vspace{-2mm}
\centering
{\includegraphics[width=0.8\columnwidth]{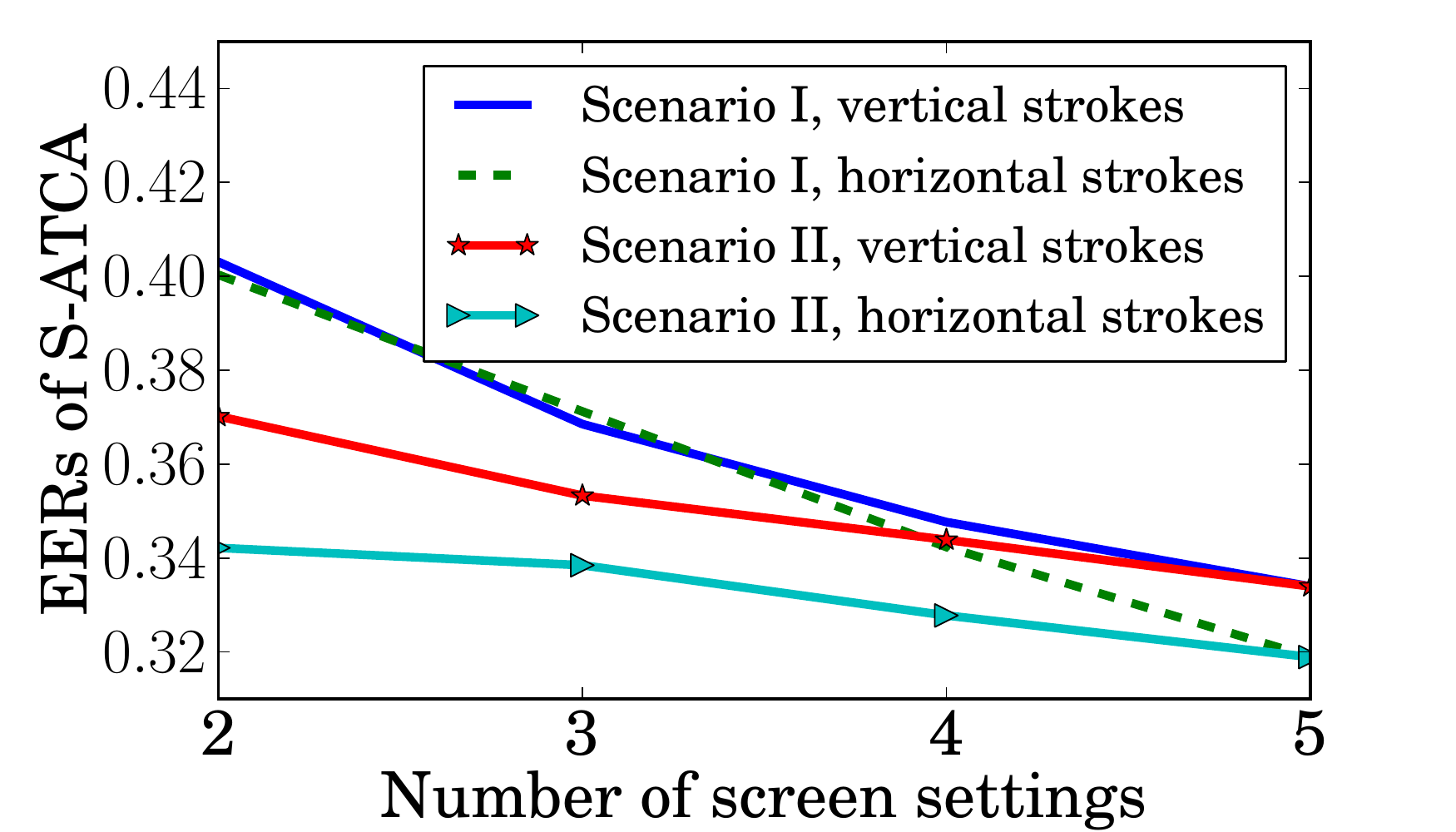}}

\caption{EERs of targeted attacks to our authentication system as a function of the number of screen settings for both scenarios and both horizontal and vertical strokes. We observe that our system can better defend against forgery attacks with more screen settings.} 
\label{impact-settings}
\vspace{-4mm}
 \end{figure}

 \myparatight{EER of authentication systems} For an authentication system and an
attack scenario (e.g., random attacks or targeted attacks), we compute an EER
via averaging the EERs of the  possible attacks to the authentication system,
 and this average EER is used to measure the resilience of
the authentication system to the attacks.

\begin{figure*}[!t]
\centering
\subfloat[Random attacks]{\includegraphics[width=0.8\columnwidth]{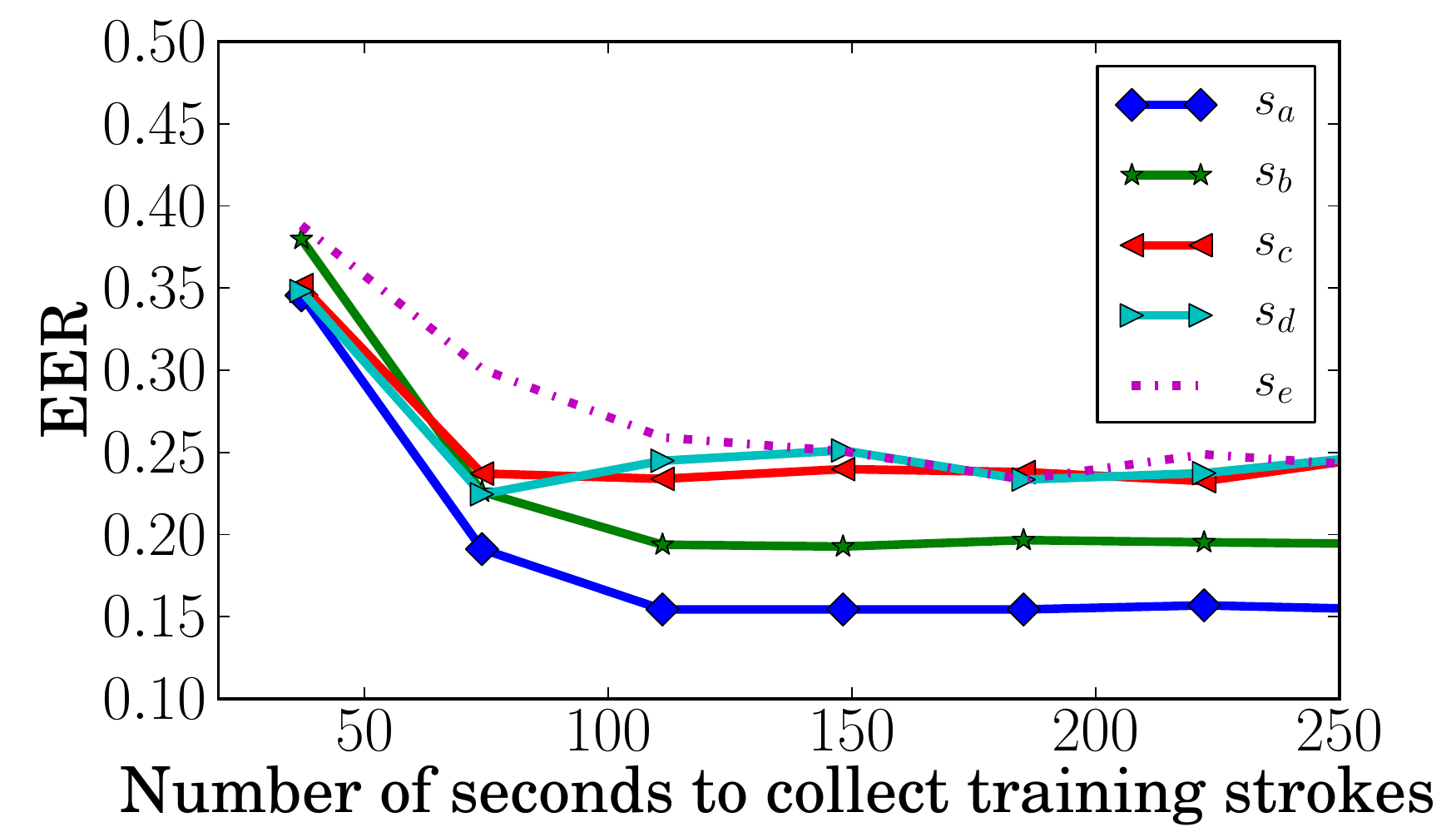}}
\subfloat[Targeted attacks]{\includegraphics[width=0.8\columnwidth]{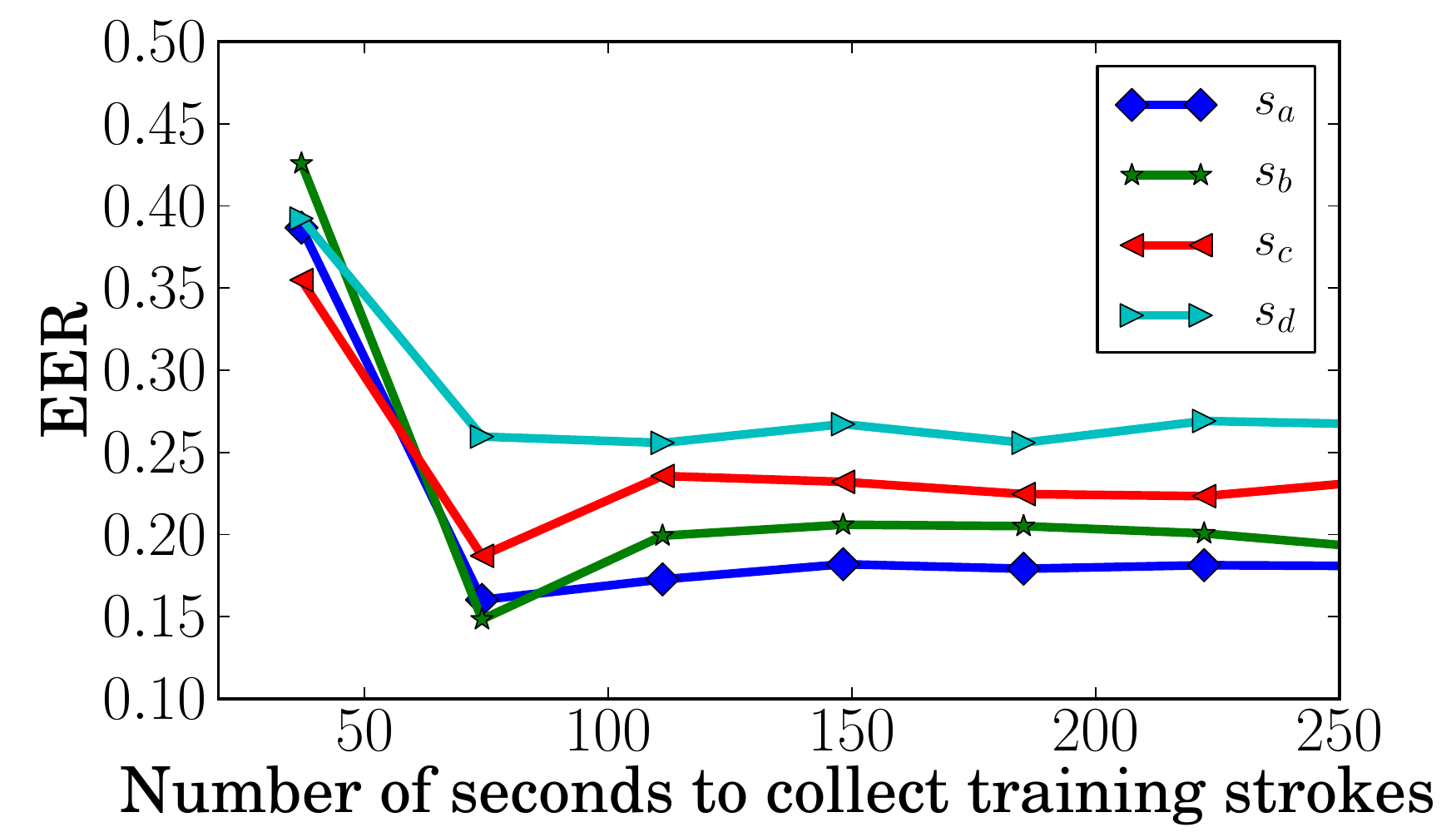}}

\caption{EERs of various attacks as a function of the time spent on collecting
training horizontal strokes. The classifier is C-ATCA-e. EERs of targeted attacks (TA) in the screen setting $s_e$ are always 0.5, and thus we do not show them in (b) to better contrast the differences of EERs in other screen settings. We find that learning our classifier is fast, i.e., the EERs are stable or slightly fluctuate after two minutes (around 30 strokes) spent on collecting training strokes.} 
\label{learning}
\vspace{-2mm}
 \end{figure*}

\myparatight{Results} Table~\ref{result-auth} shows the mean EERs over all
subjects for  each authentication system and attack scenario for horizontal strokes and vertical strokes.
 
Overall, we find that our adaptive authentication system achieves the smallest
EERs for both random attacks and targeted attacks. Specifically, the EER of adaptive authentication system 
is 2\% to 9\% smaller than those of the baseline and improved baseline authentication systems for random attacks. For targeted attacks, our improved baseline authentication system (i.e., S-Baseline-improved)  is 5\% to 6\% smaller than that of the
baseline authentication systems, and our adaptive authentication system
further decreases the EER  by 12\% for both horizontal and vertical strokes.

\subsection{Impact of the number of screen settings} We show that the EERs of our authentication system decrease when we use more appropriate screen settings. 
Towards this goal, we vary the number of screen settings and compare the corresponding EERs.

Considering the influence of the difference between two screen settings, we consider two scenarios, in which the number of screen settings increases in different fashions. 
The two scenarios are:
\begin{packeditemize}
\item {\bf Scenario I:} The new screen settings are out of the range that is covered by the existing screen settings.
Specifically, we consider two screen settings consist of $\{s_a,s_b\}$, three settings consist of $\{s_a,s_b,s_c\}$, four settings consist of $\{s_a,s_b,s_c,s_d\}$, and five settings consist of $\{s_a,s_b,s_c,s_d,s_e\}$.
\item {\bf Scenario II:} The new screen settings are in the range that is covered by the existing screen settings.
Specifically, we consider that two screen settings consist of $\{s_a,s_e\}$,
three settings consist of $\{s_a,s_c,s_e\}$, four settings consist of
$\{s_a,s_b,s_c,s_e\}$,\footnote{We also tried the other four-screen-settings $\{s_a,s_c,s_d,
s_e\}$, and we found that the two four-screen-settings achieve similar EERs.} and five settings consist of $\{s_a,s_b,s_c,s_d,s_e\}$.
\end{packeditemize}

Since EERs of random attacks are all small, we focus on targeted attacks. 
Figure~\ref{impact-settings} shows EERs of targeted attacks to our system for different number of screen settings. We observe that  our authentication system achieves smaller EERs as the number of settings increases for both horizontal strokes and vertical strokes and for both scenarios. This is because,  with less screen settings, the performance of our authentication system 
is dominated by the same-setting targeted attacks whose EERs are high. 
However, with more settings, the impact of same-setting targeted attacks is smaller,
 and the performance of our authentication system gets improved. In fact, the probability of same-setting targeted attacks
 is $\frac{1}{n}$, where $n$ is the number of
settings used.

\subsection{Learning our classifiers is fast}
To learn our classifiers for a user, we
need to collect his/her touch strokes. To study the effect of time spent on
collecting training strokes from a new user on the performance of our classifiers, 
we sample a
user and one of the five trials/folds. In the selected trial, we keep the test dataset
and attack datasets the same while increasing the positive training dataset 
(the negative training dataset is fixed).
Figure~\ref{learning} shows the EERs of various attacks as a function of time
spent on collecting positive training horizontal strokes. The classifier is C-ATCA-e. EERs
of the targeted attacks using strokes collected in the setting $s_e$ (i.e., `TA,
$s_e$') are all close to 50\%, and thus are ignored to better contrast the
differences of other EERs. 

We find that EERs converge very fast. In particular, after 2 minutes (around 30
strokes),  EERs are stable or slightly fluctuate. Moreover, after collecting
 strokes, training a classifier is finished within 1
second.

\subsection{Summary}
Our observations can be summarized as follows:
\begin{packeditemize}
\item Users can subconsciously adapt their behavior to different screen settings, 
i.e., transitions between settings do  not affect user experiences.
\item Our authentication mechanism achieves 
much smaller EERs than previous work for both random attacks and targeted attacks.  
\item Our authentication system achieves smaller EERs with more screen settings.  
\item Learning our classifiers is fast, i.e., strokes collected within two minutes are enough to stabilize  EERs.
\end{packeditemize}

\section{Discussion}
\label{discussion}
\myparatight{Training human attackers} To mimic the targeted user's touch
behavior, a human attacker needs to be trained to produce touch strokes whose
features are close to those of the targeted user. We note that Meng et
al.~\cite{tey2013can} proposed an interactive system to train a human attacker
to reproduce  \emph{keystroke dynamics} of the targeted user for a \emph{given
short password}. Specifically, they consider  features of keystroke dynamics are
constructed from 2-grams, and thus changing the keystroke timing of a character
only influences features of the local two 2-grams. For instance, suppose we have
a password with three characters ABC, changing the keystroke timing of B only
influences the features of AB and BC. Thus,  it is possible to train a human
attacker to reproduce the keystroke dynamics of a given short password via
greedily changing the keystroke timings of characters one by one. 

However,  reproducing touch strokes could be much harder than reproducing
keystroke dynamics. This is because 1) we have around 30 touch features, 2)
changing one touch point could result in changes of a few features, and 3) the
human attacker needs to learn how the targeted user would adapt to different
screen settings. Nevertheless, it is an interesting future work to explore the
possibility/impossibility of training human attackers to mimic a targeted user.

\myparatight{Fixing one screen setting to perform targeted attacks} A robot can
keep replaying touch data collected in a fixed screen setting to attack our
authentication system. The expected number of tries until the robot is using the
correct screen setting would then be  the total number of
screen settings. Once the robot gets the correct setting, the robot can use the
mobile device for a time interval during which the setting is unchanged. 

However, this attack can be blocked with a high probability by combing our
touch-based authentication with PINs. Specifically, once we detect suspicious
touch data, we ask the user to type in the backup PIN. 

\myparatight{Detecting screen settings with specialized intelligent robots} An
intelligent robot that is equipped with specialized sensors could potentially
detect the screen settings using some Artificial Intelligence (AI) algorithms,
and detecting the screen settings could enable the attacker (e.g., a friend or
spouse of the targeted user) to perform better targeted attacks.  For instance,
a robot with a camera could possibly detect the screen settings by using
computer vision algorithms to compare its raw touch data (collected via the
camera) on the screen and the movements (again, collected via the camera) of the
running application. However, the robot still needs to generate a few touch
strokes (these strokes may be from a screen setting that is different from the
one used by our authentication system) before the screen setting is detected,
during which our authentication scheme might already successfully reject the
attacker. Moreover,  it might not be easy for the attacker to get such a
specialized robot, which is true at least for now, given the current state of
AI. Therefore, we focus on robots that are commercialized and easy to get.

\myparatight{Leveraging sloppiness and jitter} Screen settings could also adjust
sloppiness and jitter other than the distortions along the X axis and the Y axis
studied in this paper.  Sloppiness controls how far the user has to move the
finger on the screen to send a movement to the applications and jitter controls
what distortions from a straight line on the screen are still considered as a
movement by the applications. It is an interesting future work to explore the
impact of sloppiness and jitter on the performance of defending against forgery
attacks in our authentication system.

\section{Conclusion and Future Work}\label{sec:conclusion}

In this work, we design a new touch-based continuous authentication system
 to defend against forgery attacks by leveraging the impact of screen 
settings on a user's touch behaviors.  First, we find that, when screen settings are discretized properly, a user's
touch behaviors in two different settings are both \emph{stable} and
\emph{sensitive}.  Second, based on these findings, we design a new authentication system called
\emph{adaptive touch-based continuous authentication}. The key idea is to randomly sample a predefined screen setting  in each time interval. The attacker cannot know the screen setting at the time of attacks. 
Third, we evaluate our system by collecting
data from 25 subjects in five screen settings. We find that  users can subconsciously 
adapt their touch behavior to different screen settings, 
i.e., transitions between settings do not interrupt users nor affect user experiences. 
 Moreover, we observe that our system
significantly outperforms previous work at defending against both random forgery
attacks and targeted forgery attacks, the registration phase of our system takes a short
period of time,  and our system can better defend  against forgery attacks with more 
screen settings.

Future work includes performing a large-scale study about our authentication system in the wild, investigating more types of screen settings,  and 
exploring more advanced attacks to touch-based authentication systems.

\balance
{
\bibliographystyle{abbrv}

}

\end{document}